\documentclass[pdflatex,sn-mathphys-num]{sn-jnl}

\usepackage{graphicx}%
\usepackage{multirow}%
\usepackage{amsmath,amssymb,amsfonts}%
\usepackage{amsthm}%
\usepackage{mathrsfs}%
\usepackage[scr=rsfs]{mathalpha}
\usepackage[title]{appendix}%
\usepackage{xcolor}%
\usepackage{textcomp}%
\usepackage{manyfoot}%
\usepackage{booktabs}%
\usepackage{algorithm}%
\usepackage{algorithmicx}%
\usepackage{algpseudocode}%
\usepackage{listings}%
\usepackage{manyfoot}%
\usepackage{subcaption}%
\usepackage{tabularx}
\usepackage{array}
\usepackage{makecell}  

\theoremstyle{thmstyleone}%

\theoremstyle{thmstyletwo}%

\theoremstyle{thmstylethree}%

\begin{document}

\title[Article Title]{Reconfigurable Intelligent Surface-Enhanced Satellite Networks: Deployment Strategies, Key Capabilities, Practical Solutions, and Future Directions}

\author[1]{\fnm{Zheng} \sur{Ziyuan}}\email{zhengziyuan2024@sjtu.edu.cn}

\author[1,2]{\fnm{Li} \sur{Xiangyu}}\email{xiangyuli@sjtu.edu.cn}

\author[3]{\fnm{Zuo} \sur{Shirui}}\email{zsr0224@bupt.edu.cn}

\author[1]{\fnm{Wan} \sur{Yu}}\email{wany35@sjtu.edu.cn}

\author[2,4]{\fnm{Shang} \sur{Bodong}}\email{bdshang@eitech.edu.cn}

\author[3]{\fnm{Jing} \sur{Wenpeng}}\email{jingwenpeng@bupt.edu.cn}

\author*[1]{\fnm{Wu} \sur{Qingqing}}\email{qingqingwu@sjtu.edu.cn}

\affil[1]{\orgdiv{Department of Electronic Engineering}, \orgname{Shanghai Jiao Tong University}, \orgaddress{ \city{Shanghai}, \postcode{200240}, \country{China}}}

\affil[2]{\orgdiv{Zhejiang Key Laboratory of Industrial Intelligence and Digital Twin}, \orgname{Eastern Institute of Technology}, \orgaddress{\city{Ningbo}, \state{Zhejiang}, \postcode{315200}, \country{China}}}

\affil[3]{\orgdiv{Beijing Key Laboratory of Network System Architecture and Convergence, Beijing Laboratory of Advanced Information Networks}, \orgname{Beijing University of Posts and Telecommunications}, \orgaddress{\city{Beijing}, \postcode{100876}, \country{China}}}

\affil[4]{\orgdiv{The State Key Laboratory of Integrated Services Networks}, \orgname{Xidian University}, \orgaddress{\city{Xi'an}, \postcode{710071}, \country{China}}}


\abstract{Satellite networks promise wide-area 6G coverage but face two persistent barriers: blockage-induced service discontinuities and increasingly stringent spectrum coexistence across satellite layers and with terrestrial systems. Reconfigurable intelligent surfaces (RISs) act as low-power programmable apertures that redirect energy without the cost and power consumption of fully active arrays. We develop a deployment-first, operations-aware view of RIS-enabled satellite networking that treats RIS as both satellite/terminal antennas and inter-satellite or space-ground relays. We show that system-level gains are governed by two unifying mechanisms: connectivity restoration via virtual line-of-sight links that preserve connectivity under blockage and mobility, and angular selectivity that reshapes interference to enlarge spectrum reuse. We further discuss practical operation under high mobility, highlighting Delay-Doppler channel acquisition, predictive beam tracking, and control designs that budget overhead and latency, and summarize hardware considerations for reliable operation in space. Finally, we outline forward-looking opportunities in the generative artificial intelligence paradigm, multifunctional RIS architectures, ubiquitous satellite integrated sensing and communication, and sustainable satellite Internet-of-Things.}

\maketitle

\section{Introduction}\label{sec1}
Satellite networks are becoming a key component in the sixth-generation (6G) system, envisioned to provide wide-area coverage and bridge connectivity gaps in remote or underserved regions, complementing terrestrial infrastructure to enable ubiquitous and seamless connectivity \cite{AlouiniSat}. However, represented by low-earth orbit (LEO) satellite constellations, satellite networks operating over long distances generally face severe path loss and limited link budgets, as well as intermittent visibility due to the terrestrial blockage. These limitations present challenges for hybrid indoor-outdoor scenarios with terrain masks and building canyons, remote Internet of things (IoT) applications with massive access and low-power terminals, and potentially integrated sensing and communication (ISAC) functionality from space. Densifying constellations or using massive active arrays can scale capacity and partially alleviate these issues. However, they come at the expense of high costs and power consumption in the deployment and operation of satellite networks. In addition, spectrum coexistence across satellite layers and terrestrial systems raises persistent interference risks, making advanced spectrum sharing solutions an urgent issue \cite{zheng2024cooperative}.

Reconfigurable intelligent surfaces (RISs), typically a planar meta-surface composed of passive phase-shift elements, can adjust the phase and amplitude of incident electromagnetic waves, thereby reshaping the reflected or refracted beams without additional active radio frequency (RF) chains \cite{QQIRS}. As a low-cost and energy-efficient method of programming the propagation environment, RIS is particularly attractive for satellite communications, where power and payload budgets are stringent and line-of-sight (LoS) links dominate \cite{Zheng_hotspot}. Operated passively or with minimal power, RISs add no thermal noise, create virtual LoS paths, and support full-duplex signal redirection. Integrated into satellite networks, RISs can dynamically steer and focus signals to bypass ground blockages and illuminate shadowed areas, improving service continuity and link budget. RISs can also facilitate spectrum sharing and interference management for satellite networks by spatially suppressing interference and creating more orthogonal propagation paths. Furthermore, RISs enable flexible deployment across multiple network sections and platforms within satellite networks.

\begin{table}[t]
\caption{List of Main Acronyms}
\centering

\begin{tabular}{ll}
\hline
\textbf{Acronym} & \textbf{Definition} \\
\hline
6G & Sixth-Generation \\
AI & Artificial Intelligence \\
BS & Base Station \\
CCSDS & Consultative Committee for Space Data Systems \\
CSI & Channel State Information \\
EIRP & Equivalent Isotropic Radiated Power \\
GenAI & Generative Artificial Intelligence \\
GEO & Geostationary-Earth Orbit \\
GNSS & Global Navigation Satellite System \\
HAPS & High-Altitude Platform Station \\
IoT & Internet of Things \\
ISAC & Integrated Sensing and Communications \\
ISL & Inter-Satellite Link \\
LEO & Low-Earth Orbit \\
LLM & Large Language Model \\
LoS & Line-of-Sight \\
MEO & Medium-Earth Orbit \\
wwWave & Millimeter Wave \\
MIMO & Multiple-Input Multiple-Output \\
NG-RAN & Next-Generation Radio Access Network \\
NTN & Non-Terrestrial Network \\
OFDM & Orthogonal Frequency Division Multiplexing \\
OISL & Optical Inter-Satellite Link \\
OTFS & Orthogonal Time Frequency Space \\
QoS & Quality of Service \\
RIS & Reconfigurable Intelligent Surface \\
RL & Reinforcement Learning \\
RF & Radio Frequency \\
SAGIN & Space-Air-Ground Integrated Network \\
SCA & Successive Convex Approximation \\
SDN & Software Defined Network \\
SDR & Semi-Definite Relaxation \\
SEE & Single-Event Effects \\
SINR & Signal-to-Interference-plus-Noise Ratio \\
SNR & Signal-to-Noise Ratio \\
THz & Terahertz \\
TID & Total Ionizing Dose \\
TTD & True-Time-Delay \\
UAV & Unmanned Aerial Vehicle \\
UE & User Equipment \\
UHF & Ultra-High Frequency \\
URLLC & Ultra-Reliable Low-Latency Communication \\
\hline
\end{tabular}
\end{table}

\subsection{Motivation and Scope}
\label{subsec:motivation_scope}

This paper is deliberately satellite-centric while being mindful of the broader non-terrestrial network (NTN)/ three-dimensional (3D) vision. Specfically, our primary motivation is that several high-impact pain points are most pronounced in satellite access and are highly aligned with what RIS can deliver, for example (i) coverage holes dominated by local blockage and low-elevation illumination that cannot be eliminated by constellation densification alone, (ii) the outdoor-indoor service gap, where continuity (rather than peak rate) is essential for many satellite-enabled applications, and (iii) operations constraints unique to satellite systems (predictable yet fast time-varying geometry, visibility windows, Doppler, stringent SWaP budgets, and for spaceborne surfaces, radiation/thermal cycling).
Accordingly, we focus on where and how RISs can be physically deployed and orchestrated in satellite networks under practical constraints, instead of re-surveying signal-processing algorithms in isolation.

Note that UAV/HAP platforms will occasionally be mentioned to indicate how specific RIS roles, e.g., geometry-agile relaying, may extend to broader 3D NTN architectures.
Nevertheless, a comprehensive treatment of NTN-wide challenges, such as aerial trajectory optimization, endurance-limited aerial relays, and multi-tier network management, is beyond the scope of this paper and remains a promising direction for future extensions built upon the satellite-first framework developed here.

\subsection{Related Surveys and Contributions}


\begin{table*}[t]
\centering
\caption{Technical scope comparison of representative survey and overview articles related to RIS-assisted satellite or NTN from 2021 to 2025.}
\label{tab:related_surveys_comparison_ris_ntn}
\small
\setlength{\tabcolsep}{2.0pt}
\renewcommand{\arraystretch}{1.12}
\begin{tabular}{|c|c|c|c|c|c|c|c|c|c|c|c|c|}
\hline
\textbf{\#} &
\rotatebox{90}{\makecell[l]{RIS-integrated\\ joint design}} &
\rotatebox{90}{\makecell[l]{Deployment\\taxonomy}} &
\rotatebox{90}{\makecell[l]{Link budget\\modeling}} &
\rotatebox{90}{\makecell[l]{Coverage\\scenarios}} &
\rotatebox{90}{\makecell[l]{Spectrum\\sharing}} &
\rotatebox{90}{\makecell[l]{Channel\\acquisition}} &
\rotatebox{90}{\makecell[l]{Resource\\allocation}} &
\rotatebox{90}{\makecell[l]{Hardware\\impairments}} &
\rotatebox{90}{\makecell[l]{Advanced RIS\\ architectures}} &
\rotatebox{90}{\makecell[l]{AI/ML}} &
\rotatebox{90}{\makecell[l]{ISAC}} &
\rotatebox{90}{\makecell[l]{IoT}} \\
\hline
\cite{Khennoufa2025MultiLayerNTN}  & \checkmark &  &  & \checkmark &  &  &  & \checkmark & \checkmark &  &  & \checkmark \\
\hline
\cite{Khan2025BeyondDiagonalRISNTN}  & \checkmark &  &  &  &  &  &  &  & \checkmark &  &  &  \\
\hline
\cite{Cao2021ConvergedRISMEC}  & \checkmark & \checkmark & \checkmark &  &  &  & \checkmark &  &  &  &  &  \\
\hline
\cite{Liu2024Empowering6GNTNIoT}  & \checkmark &  &  &  &  &  & \checkmark &  &  &  &  & \checkmark \\
\hline
\cite{Xu2021IRSSAGIN}  & \checkmark & \checkmark &  & \checkmark &  &  & \checkmark &  &  &  &  &  \\
\hline
\cite{Wu2025FLSTARRISSAGIN}  & \checkmark &  &  &  &  &  & \checkmark &  & \checkmark & \checkmark &  &  \\
\hline
\cite{Saleh2025TNNTNLocalization}  &  &  & \checkmark &  &  &  &  &  &  &  &  &  \\
\hline
\cite{Kaushik2024ISACIoT}  &  &  &  &  &  &  &  &  &  &  & \checkmark & \checkmark \\
\hline
\cite{Xiao2024LEOSAN}  &  &  &  & \checkmark &  & \checkmark &  &  &  &  &  &  \\
\hline
\cite{Luo2024LEOVLEO6G} &  & \checkmark &  & \checkmark &  &  &  &  & \checkmark & \checkmark &  &  \\
\hline
\cite{Ye2022NonterrestrialRISSProcIEEE}  & \checkmark & \checkmark & \checkmark & \checkmark &  & \checkmark & \checkmark & \checkmark & \checkmark &  &  &  \\
\hline
\cite{Wu2024DRLRISIoTMag} & \checkmark &  &  &  &  &  & \checkmark &  &  & \checkmark &  & \checkmark \\
\hline
\cite{Umer2025RISAerialNTNDRL} & \checkmark &  &  & \checkmark &  &  & \checkmark &  &  & \checkmark &  &  \\
\hline
\cite{Khan2024RIS6GNTNIoTMag} & \checkmark & \checkmark &  & \checkmark &  &  & \checkmark &  & \checkmark &  &  &  \\
\hline
\cite{Tekbiyik2022RISActionNTN} & \checkmark & \checkmark & \checkmark & \checkmark &  &  & \checkmark & \checkmark &  &  &  &  \\
\hline
\cite{Bariah2023RISSAGINIEEE_Network} & \checkmark & \checkmark &  & \checkmark &  &  & \checkmark &  &  &  &  &  \\
\hline
\cite{Toka2024RISEmpoweredLEOComMag} & \checkmark & \checkmark &  & \checkmark &  &  & \checkmark & \checkmark & \checkmark &  &  &  \\
\hline
\cite{Liu2024COMST_RIS_SATN_PDNOMA} & \checkmark & \checkmark & \checkmark & \checkmark &  & \checkmark & \checkmark & \checkmark & \checkmark & \checkmark &  &  \\
\hline
\cite{Khoshafa2025SecuredSAGIN} & \checkmark &  &  &  &  &  &  &  &  &  &  &  \\
\hline
\cite{Jamshed2024AirborneNTNRISIoT} & \checkmark &  &  & \checkmark &  &  &  &  &  &  &  & \checkmark \\
\hline
\cite{Trinh2025QuantumSAGINOpticalRIS} & \checkmark & \checkmark &  & \checkmark &  &  &  &  & \checkmark &  &  &  \\
\hline
\cite{Ramezani2023RISEnhancedINTENT} & \checkmark & \checkmark &  & \checkmark &  &  & \checkmark &  &  &  &  &  \\
\hline
\cite{Wu2025URLLC_DRL_RIS_SAGIN} & \checkmark &  &  &  &  &  & \checkmark &  &  & \checkmark &  &  \\
\hline
\makecell[l]{This\\ work} & \checkmark & \checkmark & \checkmark & \checkmark & \checkmark & \checkmark & \checkmark & \checkmark & \checkmark & \checkmark & \checkmark & \checkmark \\
\hline
\end{tabular}
\end{table*}

\begin{table*}[t]
\centering
\caption{Coverage/Spectrum-sharing oriented comparison of representative survey/overview articles related to RIS-assisted satellite/NTN (2021--2025), under a conservative marking criterion.}
\label{tab:related_surveys_10cols_deploy_cover_share}
\small
\setlength{\tabcolsep}{2.2pt}
\renewcommand{\arraystretch}{1.10}
\begin{tabular}{|c|c|c|c|c|c|c|c|c|c|c|}
\hline
\textbf{\#} &
\rotatebox{90}{\makecell[l]{Sat-mounted\\ RIS antenna}} &
\rotatebox{90}{\makecell[l]{Terminal-\\ mounted RIS}} &
\rotatebox{90}{\makecell[l]{Inter-satellite\\ RIS relay}} &
\rotatebox{90}{\makecell[l]{Space-ground\\ RIS relay}} &
\rotatebox{90}{\makecell[l]{Outdoor\\ coverage}} &
\rotatebox{90}{\makecell[l]{Indoor\\ coverage}} &
\rotatebox{90}{\makecell[l]{Mobility\\ coverage}} &
\rotatebox{90}{\makecell[l]{Intra-layer\\ sharing}} &
\rotatebox{90}{\makecell[l]{Inter-layer\\ sharing}} &
\rotatebox{90}{\makecell[l]{Cross-layer\\ sharing}} \\
\hline
\cite{Khennoufa2025MultiLayerNTN}  &  &  &  &  &  &  &  &  &  &  \\
\hline
\cite{Khan2025BeyondDiagonalRISNTN}  &  &  &  &  &  &  &  &  &  &  \\
\hline
\cite{Cao2021ConvergedRISMEC}  &  &  &  &  &  &  &  &  &  &  \\
\hline
\cite{Liu2024Empowering6GNTNIoT}  &  &  &  &  &  &  &  &  &  &  \\
\hline
\cite{Xu2021IRSSAGIN}  &  &  &  & \checkmark & \checkmark &  &  &  &  &  \\
\hline
\cite{Wu2025FLSTARRISSAGIN}  &  &  &  &  &  &  &  &  &  &  \\
\hline
\cite{Saleh2025TNNTNLocalization} &  &  &  &  &  &  &  &  &  &  \\
\hline
\cite{Kaushik2024ISACIoT}  &  &  &  &  &  &  &  &  &  &  \\
\hline
\cite{Xiao2024LEOSAN}  &  &  &  &  &  &  &  &  &  &  \\
\hline
\cite{Luo2024LEOVLEO6G} &  &  &  &  &  &  &  &  &  &  \\
\hline
\cite{Ye2022NonterrestrialRISSProcIEEE} &  &  &  & \checkmark & \checkmark &  & \checkmark &  &  &  \\
\hline
\cite{Wu2024DRLRISIoTMag} &  &  &  &  & \checkmark &  & \checkmark &  &  &  \\
\hline
\cite{Umer2025RISAerialNTNDRL} &  &  &  &  & \checkmark &  & \checkmark &  &  &  \\
\hline
\cite{Khan2024RIS6GNTNIoTMag} &  &  &  &  &  &  &  &  &  &  \\
\hline
\cite{Tekbiyik2022RISActionNTN} &  &  &  & \checkmark & \checkmark &  & \checkmark &  &  &  \\
\hline
\cite{Bariah2023RISSAGINIEEE_Network} &  &  &  & \checkmark & \checkmark &  &  &  &  &  \\
\hline
\cite{Toka2024RISEmpoweredLEOComMag} &  &  &  &  & \checkmark &  & \checkmark &  &  &  \\
\hline
\cite{Liu2024COMST_RIS_SATN_PDNOMA} &  &  &  & \checkmark & \checkmark &  &  &  &  &  \\
\hline
\cite{Khoshafa2025SecuredSAGIN} &  &  &  &  &  &  &  &  &  &  \\
\hline
\cite{Jamshed2024AirborneNTNRISIoT} &  &  &  &  & \checkmark &  & \checkmark &  &  &  \\
\hline
\cite{Trinh2025QuantumSAGINOpticalRIS} &  &  &  & \checkmark & \checkmark &  &  &  &  &  \\
\hline
\cite{Ramezani2023RISEnhancedINTENT} &  &  &  & \checkmark & \checkmark &  &  &  &  & \checkmark \\
\hline
\cite{Wu2025URLLC_DRL_RIS_SAGIN} &  &  &  &  &  &  &  &  &  &  \\
\hline
\makecell[l]{This\\ work}
& \checkmark & \checkmark & \checkmark & \checkmark
& \checkmark & \checkmark & \checkmark
& \checkmark & \checkmark & \checkmark \\
\hline
\end{tabular}
\end{table*}

RIS/IRS-enabled satellite and NTN has rapidly attracted interest, resulting in a number of surveys and overview articles in the past few years.
Representative broad overviews summarize RIS-aided NTN/space-air-ground integrated network (SAGIN) architectures, use cases, and enabling techniques from a general communication-and-networking perspective (e.g., RIS-assisted NTNs \cite{Ye2022NonterrestrialRISSProcIEEE,Tekbiyik2022RISActionNTN}, SAGIN-oriented views \cite{Xu2021IRSSAGIN,Bariah2023RISSAGINIEEE_Network}, and LEO-centric usage scenarios \cite{Toka2024RISEmpoweredLEOComMag}).
In parallel, topic-focused surveys emphasize specific dimensions such as NOMA-enabled SATN \cite{Liu2024COMST_RIS_SATN_PDNOMA}, IoT-centric NTN connectivity and learning-enabled control \cite{Liu2024Empowering6GNTNIoT,Jamshed2024AirborneNTNRISIoT,Wu2024DRLRISIoTMag,Wu2025URLLC_DRL_RIS_SAGIN,Wu2025FLSTARRISSAGIN}, beyond-diagonal/advanced RIS architectures \cite{Khan2025BeyondDiagonalRISNTN,Khan2024RIS6GNTNIoTMag}, optical/quantum SAGIN with optical RISs \cite{Trinh2025QuantumSAGINOpticalRIS}, and TN--NTN localization \cite{Saleh2025TNNTNLocalization}.
Two compact comparisons are provided in Tables~\ref{tab:related_surveys_comparison_ris_ntn} and \ref{tab:related_surveys_10cols_deploy_cover_share} to highlight their scope and emphasis.

While these surveys are contributory, Tables~\ref{tab:related_surveys_comparison_ris_ntn}--\ref{tab:related_surveys_10cols_deploy_cover_share} reveal that most existing works primarily organize the literature around technologies (e.g., NOMA/IoT/security/learning) or signal-processing methods (e.g., optimization and learning), with relatively limited emphasis on deployment roles and operations constraints (e.g., visibility windows, Doppler, control latency, SWaP, and multi-band hardware feasibility) that often determine whether an RIS-assisted design can be realized in near-term satellite systems.
In contrast, this paper is positioned as a deployment- and operations-aware synthesis: we treat RIS as an infrastructure component whose effectiveness is governed by geometry, orchestration, and hardware constraints, rather than by beamforming algorithms alone.
Relative to existing surveys, the distinctive contributions of this paper are threefold:
\begin{itemize}
  \item \textbf{RIS deployment taxonomy in satellite networks with link-budget interpretations:}
  We systematize RIS-assisted satellite networks into four representative deployment strategies, including satellite- and terminal-mounted RIS antennas, inter-satellite RIS relays, and space–ground RIS relays, and distill how their physics translates into system-level value. We provide link-budget reasoning, highlighting satellite-specific limiting factors, including deployment geometry, productive path loss, and near-/far-field regimes.
  \item \textbf{Novel perspectives on seamless coverage and spectrum coexistence:}
  We organize RIS-enabled seamless coverage via virtual LoS restoration into outdoor, indoor, and mobility scenarios, and revisit classical spectrum-sharing regimes, including intra-layer, inter-layer, and cross-layer satellite-terrestrial coexistence, through an RIS-enhanced spatial diversity and angular selectivity, clarifying what is standard versus what is incrementally enabled by RIS.
  \item \textbf{Comprehensive discussions on challenge and opportunities:}
  We map operating challenges in the high-mobility, large-scale, and heterogeneous network to pragmatic solutions, including channel acquisition and mobility management, resource allocation, and hardware imperfections. We also outline forward-looking opportunities in the generative AI paradigm, multifunctional RIS architectures, ubiquitous satellite ISAC, and sustainable satellite IoT from a feasibility viewpoint.
 \end{itemize}

\subsection{Organization}

\begin{figure}[ht]%
\centering
\includegraphics[width=\textwidth,height=0.8\textheight,keepaspectratio]{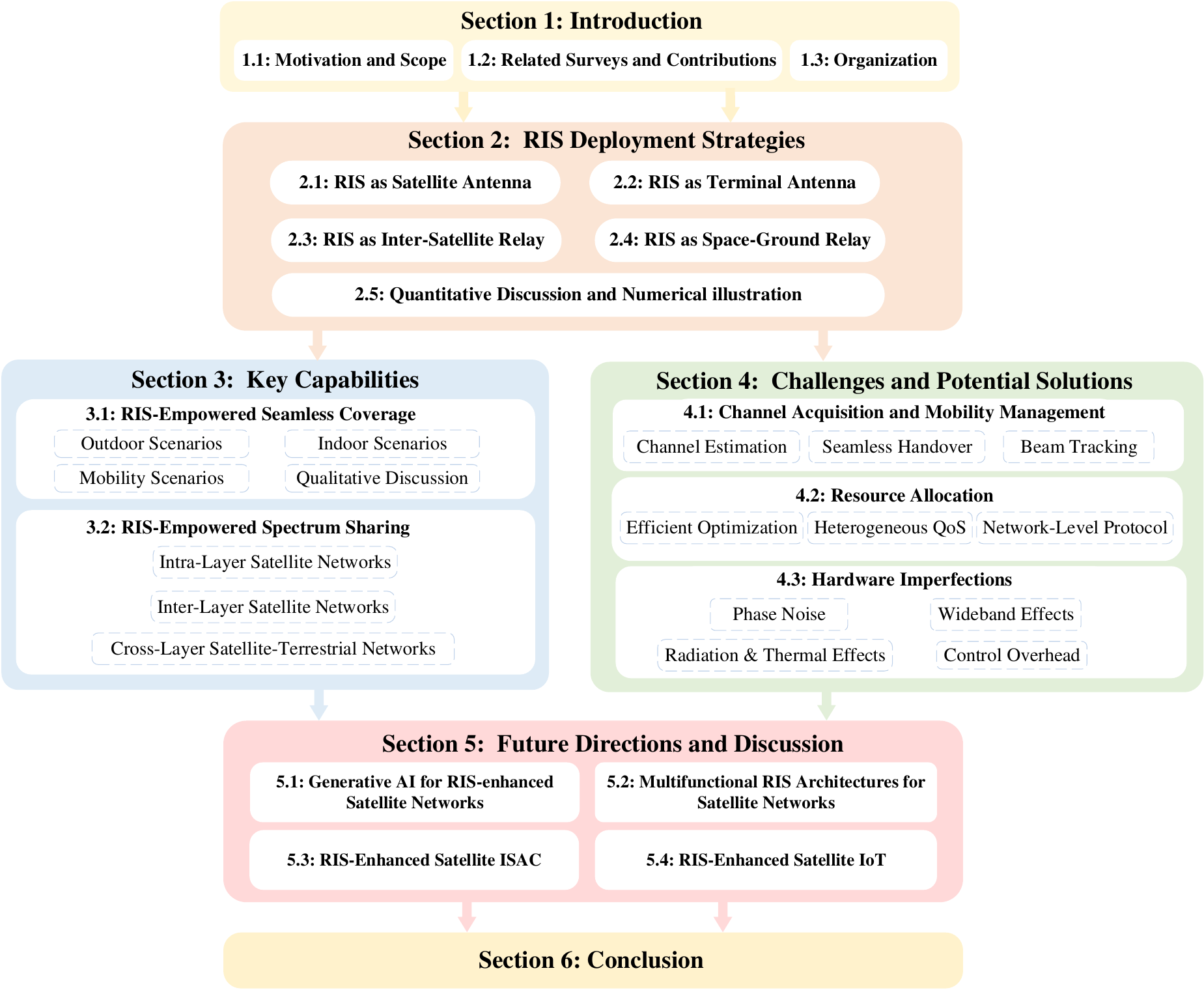}
\caption{The overall arrangement of the paper.}
\label{fig_structure}
\end{figure}

This article examines the integration of RISs into satellite networks, focusing on where RISs can be deployed, how they operate and yield benefits in key enabling scenarios, and what is required to operate them at scale. As organized in Fig.~\ref{fig_structure}, our contributions are summarized as follows. 
\begin{itemize}
\item First, we classify four typical RIS deployment strategies in satellite networks, including RIS as a satellite antenna, RIS as a terminal antenna, RIS as an inter-satellite relay, and RIS as a space–ground relay, and discuss the key benefits and deployment issues associated with each.
\item Second, we present key capabilities in RIS-enhanced satellite networks, including seamless coverage that restores LoS when geometry blocks direct paths across outdoor, indoor, and mobility use cases, and spectrum sharing that mitigates interference within or between orbital layers and across satellite–terrestrial systems.
\item Third, we address practical challenges in RIS-enhanced satellite networks by proposing pragmatic solutions, including channel acquisition and beam tracking methods, resource allocation and management designs, and countermeasures for hardware imperfections and impairments.
\item Finally, we explore and specify four future directions in RIS-enhanced satellite networks, spanning the generative artificial intelligence (AI) paradigm, multifunctional RIS architectures, ubiquitous satellite ISAC, and sustainable satellite IoT.
\end{itemize}

Collectively, these analyses position RIS-enhanced satellite networks as a compelling path toward ubiquitous, intelligent, and flexible 6G wireless networks.

%

\section{RIS Deployment Strategies in Satellite Networks}

\begin{figure}[t]
    \centering
    \includegraphics[width=3.5in]{figures/deployment.pdf}
    \caption{Four main RIS deployment schemes in satellite networks: (a) attaching an RIS to a satellite as a smart satellite antenna; (b) using an RIS at the user terminal as a lightweight antenna; (c) placing an RIS in space as an inter-satellite relay between satellites; and (d) deploying RIS as a space-ground relay between satellites and users.} 
    \label{fig:system_model_1}
\end{figure}

RISs can be deployed in many sections and nodes of satellite networks, each with different working principles, benefits, and considerations. As shown in Fig. \ref{fig:system_model_1}, this section distills four canonical deployment architectures: RIS as a satellite antenna, RIS as a terminal antenna, RIS as an inter-satellite relay, and RIS as a space–ground relay.

\subsection{RIS as Satellite Antenna}

\textbf{Working principle.} RISs can be attached to large panels onboard a satellite, such as the rear of a solar array or a dedicated deployable structure \cite{Zheng_SatIRS}. Illuminated by the onboard feed, the satellite-mounted RIS reflects and refines RF signals toward Earth, producing electronically steerable, multi-beam downlinks with reduced radio-frequency (RF) chains.

\textbf{Key Benefits.} \textit{Hardware light-weighting:} RIS, operated as a passive electromagnetic sheet, replaces hundreds of transmit modules in a phased array, slashing mass, volume, power consumption, and thermal load, while maintaining phased-array-level gain. \textit{Dynamic footprint shaping:} In operation, real-time phase shift updates allow the satellite to thin or widen beams, track hot spots, or null interference with improved performance and latency. \textit{Duplex symmetry:} As the RIS reflects signals without extra front-ends, the same surface reflects return-link signals, enabling near-full-duplex operation. 

\textbf{Deployment issues.} \textit{Aperture size vs. stowage:} A very large, multi-thousand-element aperture is desirable to compensate for path loss over hundreds-of-kilometer slant ranges, yet must fold into fairings and survive launch for feasible deployment. \textit{Field of view:} A planar RIS can lose efficiency beyond ±60°; extreme steering may require a gimbal-mounted or multiple panels, so the satellite must physically orient the surface to extend the field of view to different service areas. \textit{Radiation-resistant electronics:} The space environment affects the reliability of electronic components on the RIS. Phase shifters or varactors must hold calibration and functionality under temperature swings and radiation without on-orbit re-tuning.

\subsection{RIS as Terminal Antenna}

\textbf{Working Principle.} Integrating RISs into terminals enhances signal transmission while reducing the size, cost, and power burdens of large terminal antenna arrays, offering low power consumption, compact size, high array gain potential, and agile beam control without mechanical steering \cite{liu2021compact}\cite{hu2023securing}. 

\textbf{Key Benefits.} \textit{Conformal integration:} RIS can be embedded on vehicle roofs, unmanned aerial vehicles (UAVs) fuselages, ship decks, or equipment enclosures, while meeting tight volume, weight, and sealing constraints with ultrathin or flexible profiles, such as a reflector membrane. \textit{Cooperative sensing enhancement:} RIS can reshape the angular scattering pattern and increase the apparent mono- or bi-static radar cross section toward the sensor, thereby improving the detection and track initiation of small platforms (e.g., UAVs) under satellite-based illumination, without radiating additional power at transceivers. \textit{Multi-mode extensibility:} Reconfigurable tuning supports multibeam operation and shared aperture for transceivers with multiband extensibility. Additionally, a single panel can enable concurrent access to geostationary-earth orbit (GEO)/medium-earth orbit(MEO)/LEO satellites for path diversity and service aggregation through subarray partitioning.

\textbf{Deployment issues.} \textit{Uplink power budget:} The peak equivalent isotropic radiated power (EIRP) of the uplink is bounded by the user equipment (UE) power amplifier, the effective aperture area, and the aperture efficiency. However, most RIS are passive or semi-passive apertures, whose attainable peak transmit gain and linearity are limited relative to fully active phased arrays. Therefore, meeting stringent uplink EIRP requirements may involve a joint design of the effective aperture and hybrid active and passive amplification path. \textit{Cross-band efficiency:} Aperture efficiency and cross-band consistency of RISs are bounded by resonance bandwidth, achievable phase excursion, and insertion/ohmic loss. In Ku/Ka links, these penalties directly erode the link budget and fade margin, so terminal-mounted RISs must sustain wideband, angle-resilient phase control under rapid look-angle dynamics and high doppler, while satisfying asymmetric duplexing and circular-polarization targets.

\subsection{RIS as Inter-Satellite Relay} 

\textbf{Working principle.}
The deployment of RISs on relay satellites or other dedicated space platforms serves as a promising enhancement for terahertz (THz) inter-satellite links (ISLs), enabling passive or low-power reflection-based relaying that redirects beams between satellites when there is no direct link \cite{tekbiyik2022reconfigurableM}. This can be extended to multi-hop ISLs with multiple RIS-mounted relay satellites.

\textbf{Key benefits.}
Compared to conventional ISL schemes, which are typically dependent on RF or optical inter-satellite links (OISL) \cite{shang2025channel}, RIS-assisted relaying can provide three main benefits.
\textit{Passive relaying:} The use of RIS for ISLs eliminates the need for active radio chains. This can significantly reduce on-board power consumption and hardware complexity for the two communicating satellites.
\textit{Simultaneous redirecting:} An RIS can simultaneously support multiple ISLs with limited interference, especially for future ultra-dense satellite constellations.
\textit{Advanced steering:} By offering a broader field of reflection, RIS can mitigate the requirements for point, acquisition, and tracking in ISLs.

\textbf{Deployment issues.}
\textit{Optimal locations:} RIS placement locations, either on dedicated relay satellites or as attachments to existing satellites, exhibit trade-offs between coverage, structural feasibility, and link geometry. 
\textit{Reflection and path loss:} Reflection loss and phase shifts present key optimization problems for onboard operations of satellites, under long distances between satellites and their high mobility.
\textit{Beam squint mitigation:} Due to the use of THz bands for ISLs, which are generally high-frequency wideband scenarios, reflection-induced beam squint and split effects should be quantitatively considered for system design \cite{amodu2024technical}, especially for multi-RIS multihop topologies.

\subsection{RIS as Satellite-Ground Relay}

\textbf{Working principle.} A RIS mounted on an intermediate platform, such as high-altitude platforms (HAP), UAV, building facades, or rooftops, acts as a passive bent pipe, redirecting low-elevation satellite beams into terrestrial shadows, indoor spaces, or even underground facilities. The RIS-enhanced space-ground paths can be single-hop or multi-hop as needed.

\textbf{Key benefits.} \textit{Blockage mitigation:} RIS-aided virtual LoS paths can bypass skyscrapers, valley ridges, or underground entrances, shrinking dead zones of satellite networks without active repeaters.
\textit{Flexible placement:} Typically, satellites operate at preset orbital positions, while RIS can be flexibly deployed on demand to further expand connectivity and coverage. Both ground and airborne RISs with optimized deployment orientations and positions can fine-tune the redirection of satellite signals in target areas. \textit{Multihop cascades:} Layered space-air-ground-indoor reflections extend satellite signals to tunnels or mines, enabling ubiquitous coverage for satellite networks.

\textbf{Deployment issues.} \textit{Multi-satellite-aware scheduling:} Successive LEO satellites are visible to RIS as a space-ground relay; therefore, rapid configuration handovers and satellite-aware scheduling are required. \textit{Optimal RIS placement:} Satellite viability and path loss are a trade-off for optimal RIS placements. Ideally, RIS placement can maximize the concurrent view of satellites (e.g., the boresight of the RIS is oriented towards the sky or satellite clusters), which is a key indicator to ensure continuous service or reliable multi-connectivity; however, the path loss of the cascade channel due to the RIS radiation pattern and productive attenuation should be taken into account when placing the RIS, to maintain an adequate link budget and avoid extreme tilt angle and signal incidence angle at the RIS \cite{Zheng_deployment}.

\subsection{Quantitative Discussion and Numerical illustration}

In the following, we instantiate the link-budget model for RIS-aided satellite communication and its single-hop counterparts using representative geometries and parameter choices for the four deployment strategies in Sec.~2.

\subsubsection{Satellite link budget models under RIS far-field passive beamforming}
We adopt a widely used far-field passive beamforming operation to provide order-of-magnitude comparisons across the four deployment strategies. In general, the received power of a RIS-aided transmission, e.g., from a satellite to the RIS and then to the user, can be expressed as
\begin{align}
P_{\text{U}}
&=P_{\text{S}}
\frac{G_{\text{S}}G_{\text{U}}\lambda ^2}{16\pi ^2 d_{\text{SR}}^{\alpha}}\,
\frac{G_{\text{R}}M^2N^2d_{\text{x}}d_{\text{y}}
F\!\left( \varphi _{\text{t}},\vartheta _{\text{t}} \right)
F\!\left( \varphi _{\text{r}},\vartheta _{\text{r}} \right)\eta ^2}
{4\pi d_{\text{RU}}^{\beta}},
\label{eq:RIS_link_budget_general_rev}
\end{align}
where $P_{\text{S}}$ is the satellite transmit power, $G_{\text{S}}$ and $G_{\text{U}}$ denote the transmit and receive antenna gains at the satellite and the user terminal, respectively, and $\lambda$ is the carrier wavelength.
The distances $d_{\text{SR}}$ and $d_{\text{RU}}$ represent the link lengths between the satellite and the RIS, and between the RIS and the user, respectively, while $\alpha$ and $\beta$ denote the corresponding path-loss exponents. The factor $G_{\text{R}}$ captures the baseline aperture factor of the RIS unit cells, $M$ and $N$ are the numbers of elements along the horizontal and vertical dimensions of the RIS, $d_{\text{x}}$ and $d_{\text{y}}$ are the size of each elements, also the inter-element spacings, and hence $MNd_{\text{x}}d_{\text{y}}$ approximates the physical aperture area of the RIS.
The functions $F(\varphi_{\text{t}},\vartheta_{\text{t}})$ and $F(\varphi_{\text{r}},\vartheta_{\text{r}})$ denote the normalized array factors, or practical directional gain patterns, of the RIS in the directions of the satellite and the user, respectively, and $\eta\in(0,1]$ models the radiation efficiency of the RIS elements, including ohmic loss, phase quantization loss, and reflection loss.
The product $M^2N^2F(\varphi_{\text{t}},\vartheta_{\text{t}})F(\varphi_{\text{r}},\vartheta_{\text{r}})\eta^2$ thus compactly captures the familiar $N^2$-type power scaling of a coherently phased passive array in the desired reflecting direction.

Considering the deployment of RIS to create virtual LoS paths, equation~\eqref{eq:RIS_link_budget_general_rev} can be interpreted as the cascade of two Friis-like links, namely the satellite-RIS hop and the RIS-user hop, with the RIS acting as a passive aperture that first collects and then re-radiates the incident power. Specifically, the first fraction in~\eqref{eq:RIS_link_budget_general_rev} corresponds to the power intercepted by the RIS from the satellite; after reflection, a fraction $\eta^2$ of this collected power is re-radiated towards the user with a transmit gain proportional to the effective collecting area of the RIS $G_{\text{R}}MN d_{\text{x}}d_{\text{y}}F(\varphi_{\text{t}},\vartheta_{\text{t}})F(\varphi_{\text{r}},\vartheta_{\text{r}})$, and the second fraction in~\eqref{eq:RIS_link_budget_general_rev} can be deemed accounting for the propagation loss over the RIS-user hop. Under free-space propagation $\alpha=\beta=2$ and ideal broadside steering $F(\cdot)\approx 1$,~\eqref{eq:RIS_link_budget_general_rev} reduces to a compact double-Friis form with an overall array gain that scales proportionally to $(MN)^2$ and inversely with $d_{\text{SR}}^2 d_{\text{RU}}^2$
For the two relay-type deployments discussed in Sec.~2, namely the inter-satellite RIS relay and the space--ground RIS relay,~\eqref{eq:RIS_link_budget_general_rev} can be applied directly by setting $(d_{\text{SR}},d_{\text{RU}})$ to the corresponding satellite-RIS and RIS-user distances.
This allows us to perform simple order-of-magnitude link-budget calculations showing, for example, how a large-aperture RIS on a high-altitude platform can compensate for the additional hop and extend the coverage of a low-elevation LEO satellite, or how a RIS mounted on an auxiliary satellite can form a virtual LoS ISL in the presence of pointing constraints.
By contrast, when the RIS is integrated into the satellite or user terminal antenna, the link budget simplifies to a single-hop form  expressed as
\begin{align}
P_{\text{U}}=P_{\text{S}}\frac{G_{\text{S}}G_{\text{U}}\lambda ^2}{16\pi ^2d_{\text{SU}}^{2}}G_{\text{R}}MNd_{\text{x}}d_{\text{y}}\eta ,
\label{eq:RIS_link_budget_transceiver}
\end{align}
where we can interpret $G_{\text{R}}MN d_{\text{x}}d_{\text{y}}\eta$ as part of the effective EIRP contribution brought by the RIS. In particular, setting $G_{\text{S}}=1$ or $G_{\text{U}}=1$ indicates that RIS is the only antenna component in the satellite or terminal, respectively. These forms highlight a practical point: whereas a satellite-mounted RIS can emulate a high-gain reflector array with relatively high onboard transmit power, a terminal-mounted RIS is constrained by limited deployment space, which makes the uplink EIRP a bottleneck for high-throughput services.

\subsubsection{Numerical settings}

To complement the qualitative deployment discussion, we provide simple order-of-magnitude numerical examples based on the link-budget models in~\eqref{eq:RIS_link_budget_general_rev} and its single-hop specialization in~\eqref{eq:RIS_link_budget_transceiver}. Unless otherwise stated, we consider a Ka-band downlink at $f = 20~\mathrm{GHz}$, corresponding to a carrier wavelength of $\lambda = 15~\mathrm{mm}$, and a LEO satellite at an altitude of approximately $600~\mathrm{km}$.
The satellite transmit power is set to $P_{\mathrm{S}} = 20~\mathrm{dBW}$, and the satellite feed or array antenna has a gain of $G_{\mathrm{S}} = 30~\mathrm{dBi}$ unless the RIS fully replaces the satellite antenna.
The user terminal is modeled as an omnidirectional antenna with $G_{\mathrm{U}} = 0~\mathrm{dBi}$ in most examples, while higher-gain user antennas are also considered for comparison.
The RIS adopts half-wavelength spacing $d_{\mathrm{x}} = d_{\mathrm{y}} = \lambda/2$, a baseline gain $G_{\mathrm{R}} = 0~\mathrm{dBi}$ with ideal isotropic pattern, i.e., $F(\varphi_{\mathrm{t}},\vartheta_{\mathrm{t}}) = F(\varphi_{\mathrm{r}},\vartheta_{\mathrm{r}}) = 1$, per element, and a radiation efficiency $\eta = 0.7$.
We adopt free-space propagation with path-loss exponents $\alpha = \beta = 2$.
For signal-to-noise ratio (SNR) evaluation, we assume a system bandwidth $B = 100~\mathrm{MHz}$ and a system noise temperature $T_{\mathrm{sys}} = 500~\mathrm{K}$, which results in a noise power of $N_0 \approx -121.6~\mathrm{dBW}$.

\subsubsection{Scenario 1: Satellite-mounted RIS as a reconfigurable satellite antenna}

\begin{figure}[t]
    \centering
    \includegraphics[width=3.5in]{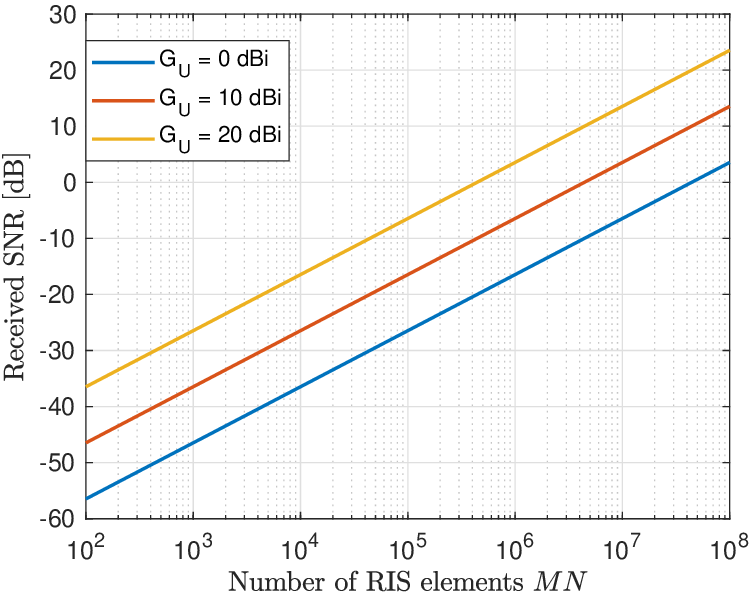}
    \caption{Received SNR versus the number of RIS elements $MN$ for different user antenna gains in the scenario of satellite-mounted RIS.} 
    \label{fig:scenario1_satRIS}
\end{figure}

In the first scenario, the RIS is mounted on the satellite and acts as the main transmit aperture while the satellite feed is modeled as an isotropic source with $G_{\mathrm{S}} = 1$.
The slant distance between the satellite and the user is fixed at $d_{\mathrm{SU}} = 600~\mathrm{km}$ (near-zenith case), and the received SNR is evaluated using the single-hop link budget ~\eqref{eq:RIS_link_budget_transceiver}.
We assume a square RIS with $M = N$ and vary the total number of elements, $MN$, from $10^2$ to $10^8$, corresponding to a progression from very small panels to extremely large apertures.
Fig.~\ref{fig:scenario1_satRIS} shows the resulting received SNR versus $MN$ for three user antenna gains $G_{\mathrm{U}} \in \{0,10,20\}~\mathrm{dBi}$. For small to moderate RIS sizes, the SNR remains well below $0~\mathrm{dB}$ even when the user terminal has a $20~\mathrm{dBi}$ antenna.
Only when the RIS aperture reaches the order of $MN \approx 10^6$ elements does the received SNR enter a comfortably positive regime for high-gain users.
The optimistic upper part of the curves (e.g., $MN \ge 10^7$) should be viewed as an asymptotic scaling illustration rather than a practical design point, since such apertures would be prohibitively large on real satellites.
Overall, this scenario highlights that satellite-mounted RISs can, in principle, emulate high-gain reflect-arrays, but achieving substantial link-margin gains at Ka-band LEO distances requires extremely large apertures or additional active amplification.

\subsubsection{Scenario 2: Terminal-mounted RIS with different platform altitudes}

\begin{figure}[t]
    \centering
    \includegraphics[width=3.5in]{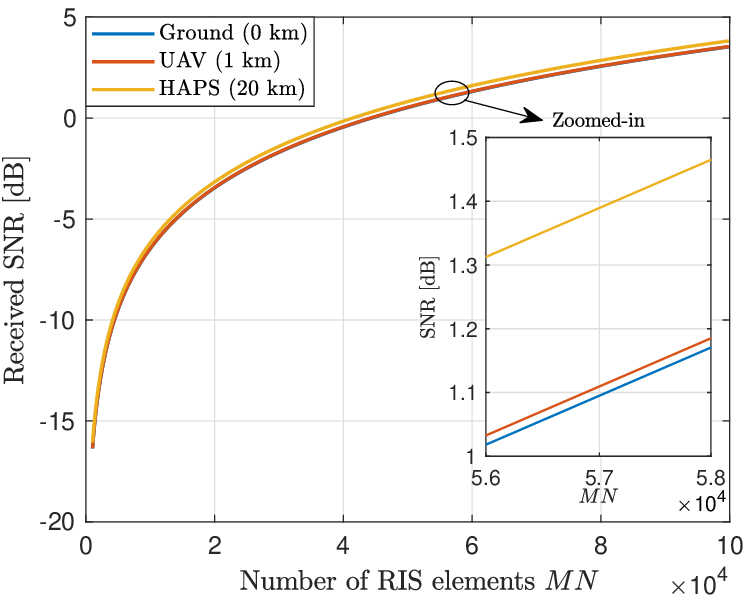}
    \caption{Received SNR versus the number of RIS elements $MN$ for different terminal altitudes in the scenario of terminal-mounted RIS.} 
    \label{fig:scenario2_termRIS}
\end{figure}

In the second scenario, the RIS is integrated into the user terminal and serves as the downlink receiving antenna.
The satellite uses its nominal $G_{\mathrm{S}} = 30~\mathrm{dBi}$ antenna, and the conventional antenna on the user side is replaced by the RIS with  $G_{\mathrm{U}} = 1$. We consider three types of terminals with different altitudes: a ground user at $0~\mathrm{km}$, a low-altitude UAV at $1~\mathrm{km}$, and a high-altitude platform station (HAPS)-like platform at $20~\mathrm{km}$.
The corresponding slant distances to a $600~\mathrm{km}$ LEO satellite are approximated as $d_{\mathrm{SU}} \approx 600$, $599$, and $580~\mathrm{km}$, respectively.
The RIS is again modeled as a square panel with $M = N$, and we vary $MN$ from $10^2$ to $10^4$ elements.
Figure~\ref{fig:scenario2_termRIS} plots the received SNR versus $MN$ for the three terminal altitudes, showing that the SNR increases monotonically with the RIS aperture size. However, the impact of terminal altitude is relatively minor: moving from a ground user to a $20~\mathrm{km}$ HAPS platform reduces the slant distance only by a few percent, resulting in sub-dB SNR improvements across the entire RIS size range, and a zoomed-in inset illustrates the small differences between the curves around $MN\approx 5.7\times10^4$.
This confirms that, for LEO satellites at several hundred kilometers, the link budget is dominated by the satellite–terminal distance rather than by modest changes in terminal altitude. 
Nevertheless, compared to UAV-mounted RISs, RISs mounted on HAPS or ground terminals can have a larger effective aperture to compensate for the large free-space losses.
These trends also foreshadow practical limitations on uplink performance: even larger user-side apertures would be required to achieve high EIRP with the constrained user transmit power.

\subsubsection{Scenario 3: Inter-satellite RIS relay}

\begin{figure}[t]
    \centering
    \includegraphics[width=3.5in]{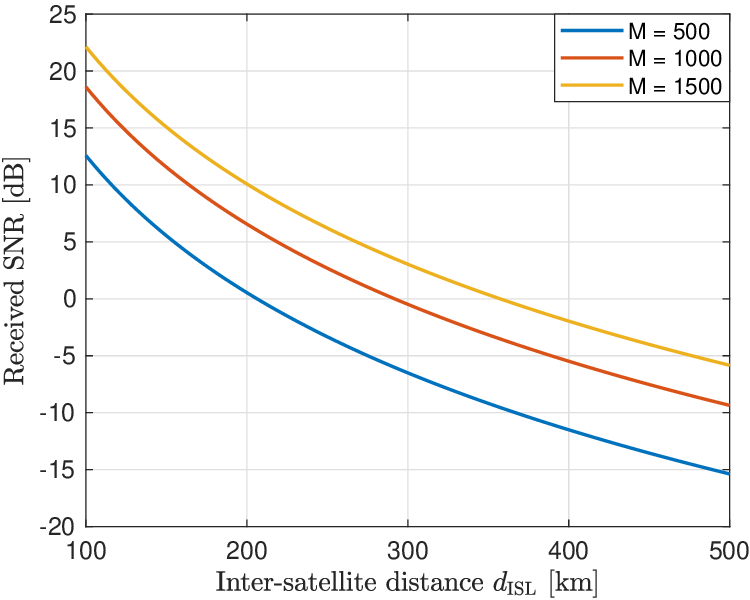}
    \caption{Received SNR versus inter-satellite distance $d_{\mathrm{ISL}}$ for different RIS sizes in the scenario of inter-satellite RIS relay.}  
    \label{fig:scenario3_ISL}
\end{figure}

The third scenario considers an inter-satellite RIS relay, in which a large passive RIS is deployed on an auxiliary satellite to establish a virtual LoS between two neighboring satellites.
We apply the double-hop model in~\eqref{eq:RIS_link_budget_general_rev} with the satellite–RIS and RIS–satellite distances both set to half of the inter-satellite distance $d_{\mathrm{ISL}}$.
We vary $d_{\mathrm{ISL}}$ from $100~\mathrm{km}$ to $500~\mathrm{km}$ and assume directive satellite antennas at both ends, i.e., $G_{\mathrm{S}} = G_{\mathrm{U}} = 30~\mathrm{dBi}$.
Fig.~\ref{fig:scenario3_ISL} shows the resulting SNR as a function of $d_{\mathrm{ISL}}$ with three RIS apertures considered: $(M,N)=(500,1000)$, $(1000,1000)$, and $(1500,1000)$.
The curves exhibit the characteristic productive double-path-loss behavior: the received SNR degrades by roughly $12~\mathrm{dB}$ when the inter-satellite distance is doubled, reflecting the fact that each hop incurs a path loss proportional to $d_{\mathrm{ISL}}^{-2}$.
For the smallest RIS, the SNR is on the order of more than ten dB at $d_{\mathrm{ISL}} = 100~\mathrm{km}$, but quickly drops below $0~\mathrm{dB}$ at $d_{\mathrm{ISL}} = 200~\mathrm{km}$.
Larger RIS apertures shift the SNR curves upward, but do not change the steep decay with distance.
These numerical examples emphasize that purely passive RIS-based inter-satellite relays can be attractive for relatively short inter-satellite distances or for modest-rate links. However, double-hop path loss sets stringent constraints on both the achievable distance and the required RIS aperture.
In many practical cases, hybrid active–passive designs or partial on-board amplification may be necessary to sustain longer-distance high-throughput ISLs.

\subsubsection{Scenario 4: Space-ground RIS relay deployed on ground infrastructures}

\begin{figure}[t]
    \centering
    \includegraphics[width=3.5in]{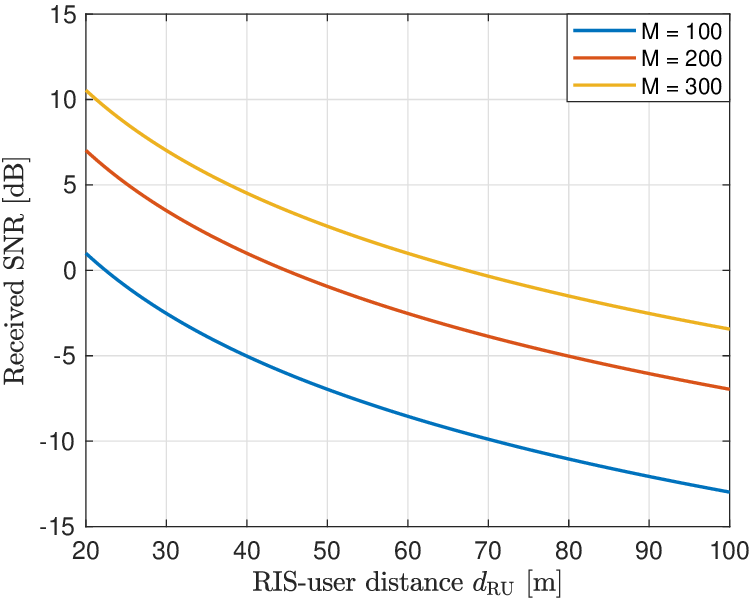}
    \caption{Received SNR versus RIS-user distance $d_{\mathrm{RU}}$ (interpretable as a coverage radius) for different RIS sizes in the scenario of space-ground RIS relay.} 
    \label{fig:scenario4_SG}
\end{figure}

In the fourth scenario, the RIS is deployed on ground infrastructures, e.g., buildings, billboards, towers, facades, or rooftops, to act as a space-ground relay between the LEO satellite and users in its vicinity.
We fix the satellite–RIS distance at $d_{\mathrm{SR}} = 600~\mathrm{km}$, consistent with a typical LEO orbit, and vary the RIS–user distance $d_{\mathrm{RU}}$ from $20~\mathrm{m}$ to $100~\mathrm{m}$, which can be interpreted as the approximate coverage radius around the RIS.
The double-hop link-budget model~\eqref{eq:RIS_link_budget_general_rev} is used with $G_{\mathrm{S}} = 30~\mathrm{dBi}$ and $G_{\mathrm{U}} = 0~\mathrm{dBi}$.
Three RIS sizes are considered: $(M,N)=(100,200)$, $(200,200)$, and $(400,200)$.
The corresponding SNR curves versus $d_{\mathrm{RU}}$ are plotted in Fig.~\ref{fig:scenario4_SG}, highlighting the trade-off between the RIS aperture and the coverage radius.
For the moderate RIS size, the SNR can be around $7~\mathrm{dB}$ at a user distance of $d_{\mathrm{RU}} = 20~\mathrm{m}$, but it drops by about $-3.5~\mathrm{dB}$ when $d_{\mathrm{RU}}$ increases by half order of magnitude to $100~\mathrm{m}$, consistent with the $d_{\mathrm{RU}}^{-2}$ dependence in the second hop.
Larger RIS apertures shift the SNR curve upward, enabling a positive SNR at tens of meters, but the SNR still degrades rapidly beyond the hundred-meter scale.
These results suggest that space-ground RIS relays are particularly well suited to create localized virtual LoS ``hotspots" over coverage radii of the order of tens of meters, especially in severe blockage environments. In contrast, kilometer-scale coverage extensions require significantly larger apertures or additional active infrastructure.

\subsubsection{Lessons learned}

The above four scenarios quantitatively corroborate the qualitative insights in Sec.~2:
\begin{enumerate}
    \item satellite-mounted RISs can emulate high-gain reflectarrays but require very large apertures at Ka-band LEO distances;
    \item terminal-mounted RISs can enhance downlink reception, yet uplink EIRP is fundamentally constrained by user power;
    \item passive inter-satellite RIS relays suffer multiplicative two-hop loss, favoring short ISLs unless apertures are very large or hybrid relaying is used; and
    \item space-ground RIS relays are best suited for localized coverage pockets whose radius scales with RIS aperture and efficiency.
\end{enumerate}

\section{Key Capabilities of RIS-Enhanced Satellite Networks}

\begin{figure}[t]
    \centering
    \includegraphics[width=5in]{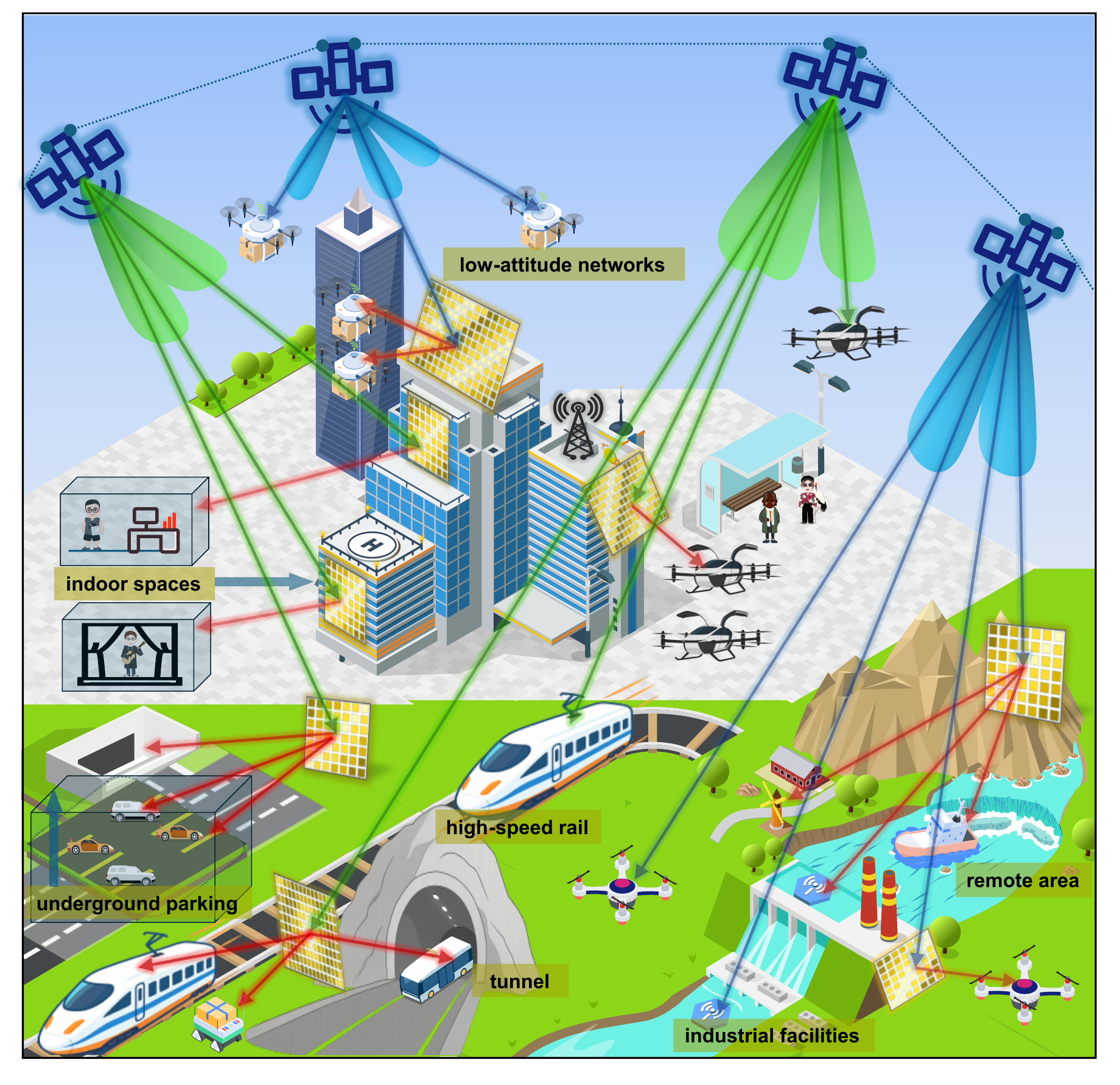}
    \caption{One of the most straightforward and compelling RIS usages in satellite networks is to improve coverage in areas where direct satellite signals are blocked or weak. The goal is to achieve seamless service coverage without dead zones in urban areas, mountainous terrain, inside buildings, and other challenging environments by utilizing RISs to redirect satellite signals on demand.} 
    \label{fig:system_model_2}
\end{figure}

With flexible deployment strategies, RIS-enhanced satellite networks can unlock a range of improvements. As depicted in Fig. \ref{fig:system_model_2}, this section discusses two key application scenarios, seamless coverage and spectrum sharing, and their performance enhancements, where the integration of RIS in satellite networks is particularly promising.

\subsection{Seamless Coverage}

A primary advantage of applying RIS in satellite networks is to fill in coverage holes caused by geometric blockage or low-elevation illumination. With proper deployment and transmission designs, RIS establishes controllable virtual LoS links that complement the direct satellite–user path. This subsection outlines representative cases for outdoor, indoor, and mobility scenarios, highlighting practical configuration choices that determine signal coverage and continuity.

\subsubsection{Seamless Coverage for Outdoor Scenarios}

In urban canyons, facades and rooftops at suitable heights can host RIS panels that redirect incident beams along streets or into courtyards when the direct low-elevation path is occluded \cite{Zheng_coverage}. In mountainous or sparsely populated regions, RIS on ridgelines or towers can illuminate valleys or reverse slopes without the need for active repeaters or backhaul. Deployment planning prioritizes locations with a long, predictable satellite–RIS visibility window and an orientation that yields a sufficiently high reflected elevation angle to avoid secondary blockage and excessive ground multipath fading. 

Using intermediate reflectors, including RIS, to bypass blockage is a known coverage concept. Here, we do not claim the concept itself is new; instead, we emphasize a satellite-specific, often under-articulated point: increasing constellation density at certain orbital altitudes generally improves pass frequency, availability, and median elevation. However, it does not eliminate blockages limited by terrestrial geometry or near-ground geometric blockages, such as street canyons or terrain masks, where many sites still experience intermittent or low-elevation illumination that is severely shadowed by local obstacles. Therefore, for coverage gaps dominated by local occlusions, the RIS is positioned as a more cost-effective complementary and relocatable infrastructure lever that restores an indirect LoS under explicit geometry.

\subsubsection{Seamless Coverage for Indoor Scenarios}

Indoor satellite coverage is traditionally considered impractical due to penetration losses from walls, glazings, and floors, and most existing satellite discussions focus on outdoor blockage. However, many satellite services, such as intelligent transportation and industrial robot networks, routinely cross the boundary between indoor and outdoor environments. We therefore synthesize the indoor coverage problem as an outdoor-indoor continuity gap relevant to emerging satellite services that cross building boundaries.

RIS can extend satellite communication coverage by creating controllable indirect paths to various indoor scenarios across facades or entrances, enabling continuity from outdoor areas into indoor spaces when the geometry and link budget permit. A single-hop solution is effective when the exterior-to-interior path includes a viable portal, such as a window, glass wall, or entrance hall. For example, a transmissive window-mounted RIS couples energy directly inside, working similarly to the outdoor cases. However, deep-indoor or underground spaces without a portal may require cascaded multi-hop redirection. Designs that aim to minimize the hop count, bound the additional path length, and maintain beam alignment are crucial in avoiding excessive productive path loss. Multi-hop RIS-aided schemes may yield modest rates, but are suitable for sensors or emergency communications, and can cooperate with short-range indoor active distribution equipment to further improve throughput.

\subsubsection{Seamless Coverage for Mobility Scenarios}

For users with high mobility, such as vehicles, trains, UAVs, and aircraft, traversing different coverage zones, RIS supports satellite communications to maintain connectivity and continuity, thereby reducing outage probability and SNR fluctuation by reconfiguring reflection states along expected routes. Note that RIS-assisted mobility has been explored in terrestrial UAV/HAPS settings. Here, we reorganize the discussion from a satellite-network viewpoint, where deterministic satellite ephemeris and predictable blockage maps enable proactive RIS scheduling and handover assistance. The focus is on system-level continuity mechanisms, including phemeris-aware triggering, codebook-based configuration with bounded overhead, and geometry-driven selection among facade/rooftop/UAV/HAP RIS options, rather than claiming a new RIS concept for mobility itself.

For example, as an aerial vehicle approaches an urban boundary from the open sky, the network schedules a facade or rooftop RIS based on satellite ephemeris, blockage maps, and the user's position. When the direct satellite link degrades, the reflected path maintains the link budget until a favorable path or another satellite becomes available. RIS mounted on a UAV/HAP can further enhance geometric agility by repositioning to preserve favorable bistatic angles for both satellite and mobility users, thus increasing the LoS probability and reducing outages. To keep control overhead bounded in mobility scenarios, element clustering and precomputed phase codebooks indexed by satellite pass, user sector, and platform pose are preferred. Compared to having multiple satellites or terrestrial base stations (BSs) handle handover for high-mobility users, an RIS can locally manage link continuity as long as it remains in favorable positions with respect to both sides. 

\subsubsection{Qualitative Discussion}

\begin{figure}[t]
    \centering
    \includegraphics[width=4.5in]{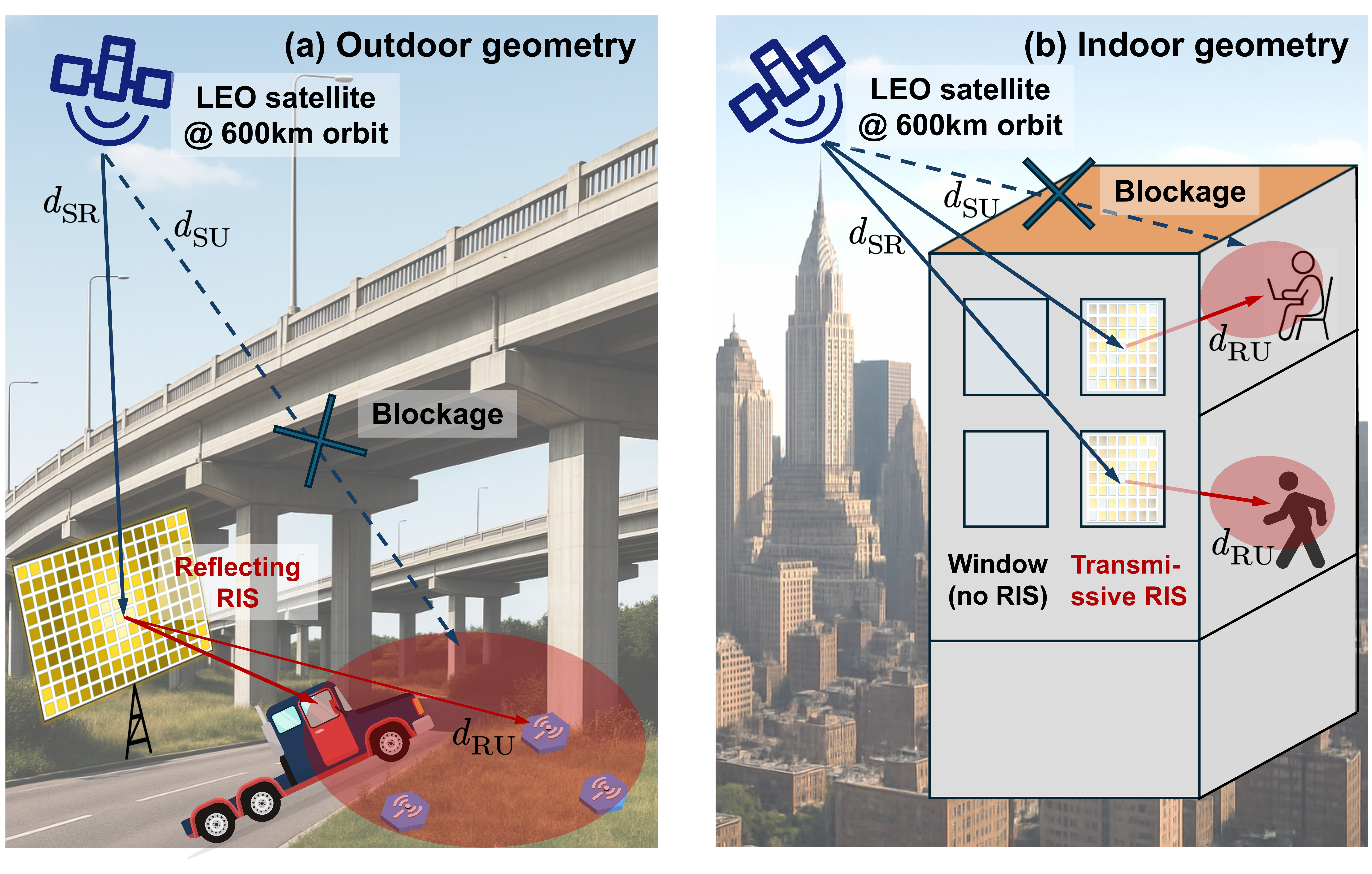}
    \caption{Representive outdoor and indoor geometries of RIS-aided satellite coverage enhancement scenarios.} 
    \label{fig:geom_seamless}
\end{figure}

\textbf{Near-field effects and multi-RIS multi-hop power limits.}
The link budgets in \eqref{eq:RIS_link_budget_general_rev} and \eqref{eq:RIS_link_budget_transceiver} rely on a far-field beamforming assumption. In seamless coverage, the satellite-RIS hop is typically far-field due to the LEO slant range, whereas the RIS-user hop may fall within the Fresnel region when users are tens of meters from meter-scale RISs. For a point-like user, near-field effects mainly shift the optimal phase profile toward spherical-wave focusing rather than plane-wave steering and do not severely invalidate the order-of-magnitude scaling, unless the distance from each RIS element to the user is significantly different, which means the user is very close to the RIS and benefits little from limited coverage extension. Thus, we retain \eqref{eq:RIS_link_budget_general_rev} and interpret $F(\cdot)$ as a generalized focusing/steering gain.
For cascaded multi-hop deployments (RIS-RIS links), however, naively multiplying far-field Friis-like terms may substantially overestimate the received power and even lead to an unphysical infinite-range implication because the next RIS cannot arbitrarily collect all power radiated by the previous finite aperture at a significant near-field due to diffraction and spillover. A conservative way to enforce energy conservation is to introduce an inter-aperture coupling factor $\kappa_{i\rightarrow j}\in(0,1)$ for each RIS-RIS hop, preventing the unphysical implication that the hop power gain could exceed one. Consequently, a multi-RIS path cannot be lossless; its end-to-end power transfer is upper-bounded by the far-field case and, in practice, decays much faster due to $\kappa_{i\rightarrow j}<1$ caused by diffraction/spillover and misalignment. This highlights that multi-hop RISs are primarily beneficial for geometric feasibility, e.g., to bypass blockages or ensure control signal coverage, rather than for unlimited range extension with high data-rate transmission.

\textbf{Representive geometry and rule-of-thumb condition.}
To clarify when a virtual LoS path through a relaying RIS remains competitive, we consider the simple geometry in Fig.~\ref{fig:geom_seamless}: (a) a RIS deployed around the road and outdoor users located under the bridges; (b) a RIS deployed on a building facade and an indoor user located at a horizontal distance $d_{\mathrm{RU}}$ from the RIS. The satellite-RIS distance is denoted by $d_{\mathrm{SR}}$, and $d_{\mathrm{SU}}$ denotes the satellite-user distance in the hypothetical unobstructed LoS case.
Using the model in~\eqref{eq:RIS_link_budget_general_rev} and the standard single-hop LoS link budget, the ratio between the received powers with a virtual LoS via the RIS and with a direct LoS can be approximated as
\begin{align}
\frac{P_{\mathrm{U}}^{\mathrm{(vLoS)}}}{P_{\mathrm{U}}^{\mathrm{(LoS)}}}
&\propto
\frac{M^2N^2d_x^2d_y^2\eta ^2d_{\text{SU}}^{2}}{4\pi d_{\text{SR}}^{2}d_{\text{RU}}^{2}}.
\end{align}
Imposing that the virtual LoS should be at most $\Delta$~dB weaker than an unobstructed LoS, i.e.,
$P_{\mathrm{U}}^{\mathrm{(vLoS)}} / P_{\mathrm{U}}^{\mathrm{(LoS)}} \ge 10^{-\Delta/10}$, yields the rule-of-thumb condition
\begin{align}
M^2N^2d_xd_y\ge \frac{4\pi d_{\text{SR}}^{2}d_{\text{RU}}^{2}}{\eta ^2d_{\text{SU}}^{2}}10^{-\Delta /10}
\label{eq:RIS_aperture_rule}
\end{align}
Define $A_{\mathrm{RIS}}=MNd_xd_y$ as the size of RIS effective aperture, we have
\begin{align}
A_{\text{RIS}}\ge \frac{2\sqrt{\pi}d_{\text{SR}}^{}d_{\text{RU}}^{}\sqrt{d_xd_y}}{\eta d_{\text{SU}}^{}}10^{-\Delta /20}
\end{align}
For a RIS-aided LEO scenario with $d_{\mathrm{SU}}\approx d_{\mathrm{SR}}\approx 600~\mathrm{km}$, $d_x=d_y=\lambda/2$ and $\eta = 0.7$, requiring the virtual LoS to be within $\Delta=0,3,10$~dB of an unobstructed LoS, respectively, lead to
\begin{subequations}
\begin{align}
A_{\text{RIS}}^{0\text{dB}}=2.53d_{\text{RU}}^{}\lambda, 
\\
A_{\text{RIS}}^{-3\text{dB}}=1.79d_{\text{RU}}^{}\lambda, 
\\
A_{\text{RIS}}^{-10\text{dB}}=0.80d_{\text{RU}}^{}\lambda. 
\label{eq:RIS_aperture_thumb_rule}
\end{align}
\end{subequations}
In other words, for a representative Ka-band communication with $f=20~\mathrm{GHz}$ ($\lambda\approx 0.015~\mathrm{m}$), a facade-mounted RIS with an aperture of about $1~\mathrm{m}^2$ can keep the virtual LoS within roughly $0$, $-3$, and $-10$~dB of an unobstructed LoS for users up to $26$, $37$, and $83$ $\mathrm{m}$ away, respectively, under the optimistic free-space and ideal-steering abstraction. At $20~\mathrm{GHz}$ with half-wavelength spacing, an aperture of $A_{\mathrm{RIS}}=1~\mathrm{m}^2$ corresponds to approximately a panel of about $133\times 133$ elements. This simple geometry and the rule-of-thumb in~\eqref{eq:RIS_aperture_thumb_rule} indicate that virtual LoS paths via large RISs are competitive for users located within tens of meters of the RIS deployment location.
The above rule focuses on a single RIS creating a virtual LoS. In deeper indoor/underground cases, multiple RISs or additional reflections may be used, but each hop introduces (i) efficiency loss and (ii) finite-aperture coupling loss to overcome indoor penetration loss without RIS. At Ku/Ka bands, typical values of penetration loss can reach $L_{\mathrm{indoor}}\approx 15-30$~dB through the wall or floor. Low-rate IoT devices, which often target SNRs around $0$--$5~\mathrm{dB}$ with narrowband or robust modulation schemes, can tolerate substantially lower received power and may remain serviceable even when the virtual LoS suffers an additional $20~\mathrm{dB}$ of indoor loss, which can be achieved by approximately three cascaded RISs with each $-6$ dB configuration, or two cascased RISs with each $-10$ dB configuration according to (6). Therefore, virtual LoS paths created by multiple RISs are well-suited for quasi-LoS IoT coverage and for users near windows or in lightly obstructed areas. In contrast, deep indoor broadband coverage is more realistically achieved through dedicated indoor infrastructure or hybrid active-passive architectures. 
\begin{table}[t]
  \centering
  \small
  \caption{A optimistic Rule-of-thumb coverage radii for single RISs creating virtual LoS paths for a representative LEO downlink.
  Distances refer to the RIS-user separation $d_{\mathrm{RU}}$ at which the virtual-LoS power equals ($0$~dB), is comparable to ($-3$~dB), or is one-tenth of ($-10$~dB) the unobstructed LoS power.
  Element counts are approximate for half-wavelength spacing and square panels.}
  \label{tab:ris_rule_of_thumb}
  \begin{tabular}{|l|l|c|c|c|}
    \hline
    Band & $A_{\mathrm{RIS}}$ [m$^2$] & Approx.\ $M\times N$ &
    $d_{\mathrm{RU}}$ [m] for $P_{\mathrm{vLoS}}/P_{\mathrm{LoS}}= 0, -3, -10$ dB \\
    \hline
    Ku (12 GHz) & 0.5 &
    $\approx 56\times 56$ &
    $7.90 \;/\; 11.16 \;/\; 24.98$ \\
    \hline
    Ku (12 GHz) & 1 &
    $\approx 80\times 80$ &
    $15.80 \;/\; 22.31 \;/\; 49.96$ \\
    \hline
    Ku (12 GHz) & 2 &
    $\approx 113\times 113$ &
    $31.59 \;/\; 44.63 \;/\; 99.91$ \\
    \hline
    Ka (20 GHz) & 0.5 &
    $\approx 94\times 94$ &
    $13.16 \;/\; 18.60 \;/\; 41.63$ \\
    \hline
    Ka (20 GHz) & 1 &
    $\approx 133\times 133$ &
    $26.33 \;/\; 37.19 \;/\; 83.26$ \\
    \hline
    Ka (20 GHz) & 2 &
    $\approx 189\times 189$ &
    $52.66 \;/\; 74.38 \;/\; 166.52$ \\  
    \hline
  \end{tabular}
\end{table}
To obtain intuitive numbers, we present a table of coverage radii for Ku/Ka band, instantiating a geometry with $d_{\mathrm{SR}}\approx d_{\mathrm{SU}}\approx 600$~km and $\eta=\eta_{\mathrm{ap}}=0.7$. We consider Ku and Ka carriers at $f=12$ and $20$~GHz, with half-wavelength spacing so that $MN\approx 4A_{\mathrm{RIS}}/\lambda^2$.

\subsection{Spectrum Sharing}

Due to spectrum scarcity, dense LEO deployments, and cross-layer coexistence requirements, spectrum sharing has long been a central issue in satellite and satellite–terrestrial integrated networks. 
Existing studies have investigated spectrum reuse and interference management within single orbital layer \cite{park2022tractable}, across multiple satellite layers \cite{okati2023stochastic}, and between satellite and terrestrial networks \cite{shang2024multi}.

Based on the above paradigms, this subsection revisits these well-studied scenarios from an RIS-enhanced satellite perspective. Specifically, RIS provides an additional, largely passive and geometry-aware degree of freedom, and enables angular selectivity and spatial interference shaping that complements conventional active beamforming and scheduling.

\begin{figure}[t]
    \centering
    \includegraphics[width=5.2in]{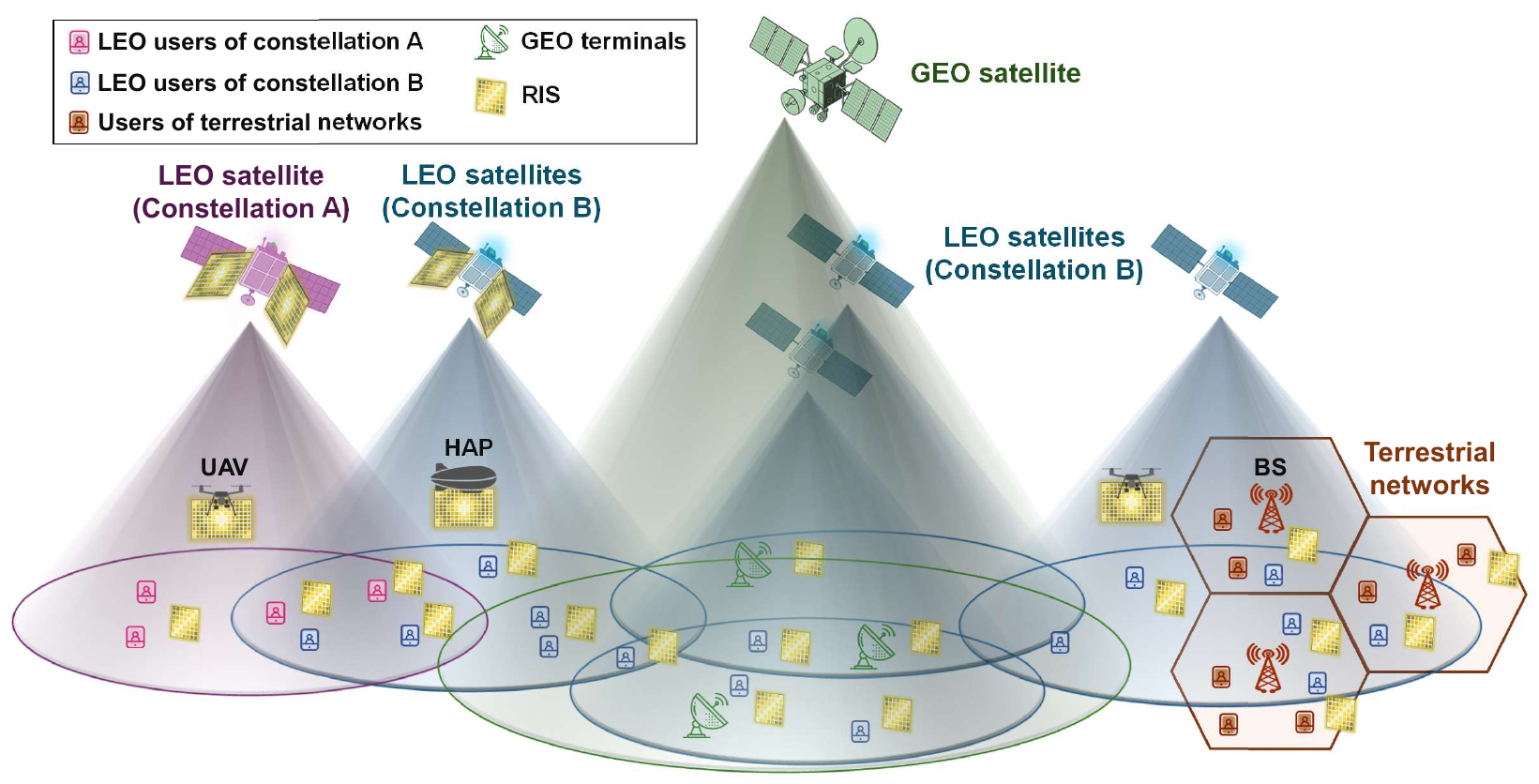}
    \caption{RIS-empowered spectrum sharing and interference management across three representative deployment scenarios for satellite networks, i.e., intra-layer satellite networks, inter-layer satellite networks, and cross-layer satellite-terrestrial networks.} 
    \label{fig:system_model_3}
\end{figure}

As shown in Fig. \ref{fig:system_model_3}, RIS can be deployed across different network layers to facilitate spectrum in satellite networks.
This section organizes RIS-empowered spectrum sharing and interference management into three representative spectrum sharing scenarios, i.e., intra-layer satellite networks, inter-layer satellite networks, and cross-layer satellite-terrestrial networks. 
Accordingly, we further clarify where RIS offers incremental leverage, how its role differs across regimes, and what interference characteristics are most amenable to RIS-based control.

\subsubsection{Intra-Layer Satellite Networks}

Intra-layer spectrum sharing among multiple satellites at the same orbital altitude is a well-known challenge in dense constellations, such as LEO mega constellations \cite{li2025downlink}. This is mainly due to the overlap of the footprints and the limited angular separation between neighboring satellites.
Conventional approaches address this issue through beam scheduling, frequency reuse planning, side-lobe suppression, and power control at the satellite payload.
From an RIS-enhanced perspective, spectrum sharing can be further improved by introducing an additional, largely passive spatial degree of freedom that enables angular interference shaping at selected ground or in-orbit locations.

Ground-deployed RIS can be placed in urban areas, especially hotspot regions, to redirect both desired and interference signals. It can not only dynamically null beams from adjacent interfering satellites to prevent direct overlap and mitigate interference, but also reflect desired signals toward the target users along angles without or with limited interference \cite{Zheng_hotspot}.
Alternatively, in-orbit RIS is deployed as transmit antennas or inter-satellite relays. It can steer outgoing and side-lobe signals to reduce interference with neighboring satellites, with only phase configuration rather than complex RF chains.

\subsubsection{Inter-Layer Satellite Networks}

Inter-layer spectrum sharing between LEO, MEO, and GEO satellite systems is a classical coexistence problem in multi-layer satellite architectures \cite{voicu2024handover}, which may reuse the same or overlapped frequency bands to support their independent services.
Existing studies have extensively analyzed inter-layer interference arising from disparities in altitudes, beam footprint size, and elevation-angle distributions \cite{li2025enriched,abudureheman2025resource,han2025demand}.
In this context, RIS does not redefine inter-layer sharing mechanisms, but offers a geometry-aware means to enhance angular separation and selectively reshape interference paths across orbital layers.

Due to frequency band reuse, these footprint overlaps will also introduce cross-layer interference, which originates from fundamental disparities in elevation angles, beam sizes, and transmission power levels \cite{radhakrishnan2016survey}. 
For example, GEO satellites, which are often observed at lower and more stable elevation angles, emit wide-area beams with high transmit power.
LEO satellites produce narrow and high-resolution beams with lower transmit power at different rapidly varying elevation angles due to their higher mobility at lower orbital altitudes \cite{prol2022position}.

RIS can be deployed on the satellites or on the ground to mitigate such interference by enhancing angular separation and beam selectivity between orbital layers \cite{zheng2024cooperative}. 
RIS mounted on LEO relay satellites or other high-altitude platforms in space can redirect interference signals from MEO/GEO satellites, thereby minimizing the impact of interference on LEO satellites.
Furthermore, ground-based RIS can be configured to selectively reflect signals toward or away from specific elevation sectors, creating angular separation across orbital layers \cite{khoshafa2025ris}.

\subsubsection{Cross-Layer Satellite-Terrestrial Networks}
In satellite-terrestrial integrated networks, spectrum sharing between satellite networks and terrestrial networks presents significant interference challenges \cite{li2025advancing}, especially when both operate over partially overlapping frequency bands, such as the S/L/Ka bands or emerging millimeter wave (mmWave)/THz shared spectrum scenarios.
This will lead to potential interference for both downlink and uplink transmissions \cite{kim2024spectrum}, considering the existence of main lobes and side lobes of both satellites and BSs.

Conventional solutions primarily rely on exclusion zones \cite{shang2025spectrum_c}, power constraints, coordinated scheduling, and regulatory protection mechanisms to manage cross-layer interference \cite{shang2025spectrum_m}.
RIS introduces a practical and complementary approach to mitigate such cross-layer interference by selectively controlling the propagation of electromagnetic waves, which can locally decouple satellite and terrestrial links without modifying transmitter-side protocols.

In downlink scenarios, ground RIS can be deployed on rooftops or near cell edges to steer satellite downlink interference signals away from sensitive terrestrial network users. 
For uplink cases, RIS can guide the signals of terrestrial users to avoid direct interference with satellite terminals, gateways, or payloads. 
Furthermore, airborne RIS mounted on UAVs or high-altitude platforms can dynamically manage air-ground interference to provide adaptive spatial filtering and beam shaping.

\text{}

The originality of RIS-enhanced spectrum sharing lies not only in redefining sharing paradigms, but also in recasting interference management as a geometry- and angle-controllable problem, which is particularly well aligned with the deterministic mobility and large-scale structure of satellite networks.

\section{Challenges and Practical Solutions in RIS-Enhanced Satellite Networks}

\subsection{Channel Acquisition and Mobility Management}

RIS-enhanced satellite networks face significant challenges in channel acquisition and beam tracking due to satellite mobility, long propagation delays, and the dimensionality introduced by the RIS. The precise acquisition and alignment of channel state information (CSI) is critical for coverage, link quality, and interference management. RIS-enhanced satellite networks face stringent channel acquisition and beam-tracking requirements due to satellite mobility, long propagation delays, and the RIS-induced cascaded, high-dimensional channel. Timely and accurate CSI acquisition is therefore critical to coverage, link quality, and interference management.

Accordingly, we first recall that orthogonal time frequency space (OTFS) maps information symbols to the delay--Doppler domain and yields an time-invariant effective channel representation that coherently captures the channel’s Delay-Doppler diversity~\cite{Hadani2017OTFSWCNC}. 
To handle the resulting Delay--Doppler-domain interference structure, \cite{Raviteja2018TWC_OTFS_IC} derives an explicit OTFS input--output relation and proposes an iterative message-passing detector with interference cancellation that is robust to doppler spread. For receiver-side acquisition in delay--Doppler channels, \cite{Raviteja2019TVT_OTFS_CE} develops embedded pilot-aided channel estimation schemes that estimate the Delay-Doppler channel response needed for OTFS detection. Targeting LEO dynamics, \cite{Shen2022JSAC_MassiveMIMOOTFS} studies grant-free random access with massive multiple-input multiple-output (MIMO)-OTFS for LEO satellite links and develops the associated signal model and detection pipeline under large delay and doppler. 
Extending OTFS to RIS cascades, \cite{Bhat2023CL_RISOTFS_Fractional} derives the end-to-end Delay-Doppler input--output relation for RIS-aided OTFS with fractional doppler and reports performance advantages over RIS-aided orthogonal frequency division multiplexing (OFDM) baselines. 
Building on these developments and the broader RIS-assisted OTFS taxonomy and design considerations summarized in \cite{Tao2025OJVT_RIS_OTFS_Survey}, we synthesize the literature to draw operational guidance for RIS-assisted satellite operation, highlighting how cascaded satellite--RIS geometry shapes the effective Delay-Doppler coherence and how training overhead and CSI freshness should align with RIS reconfiguration, handover and beam-tracking timescales.

\subsubsection{Channel Estimation}

The high-speed movement of LEO satellites induces doppler shifts and short channel coherence times. RIS exacerbates the complexity of channel estimation by adding cascaded satellite-RIS and RIS-user links. Traditional pilot-based methods, e.g., OFDM, struggle to meet real-time demands in dynamic environments with high-dimensional channels. This is because OFDM requires frequent and dense pilot insertion to track rapid time-frequency domain channel variations, leading to excessive overhead and computational burden in large-scale systems.

\begin{figure}[htbp]
    \centering
    \begin{subfigure}{0.45\textwidth}
        \centering
        \includegraphics[width=\textwidth]{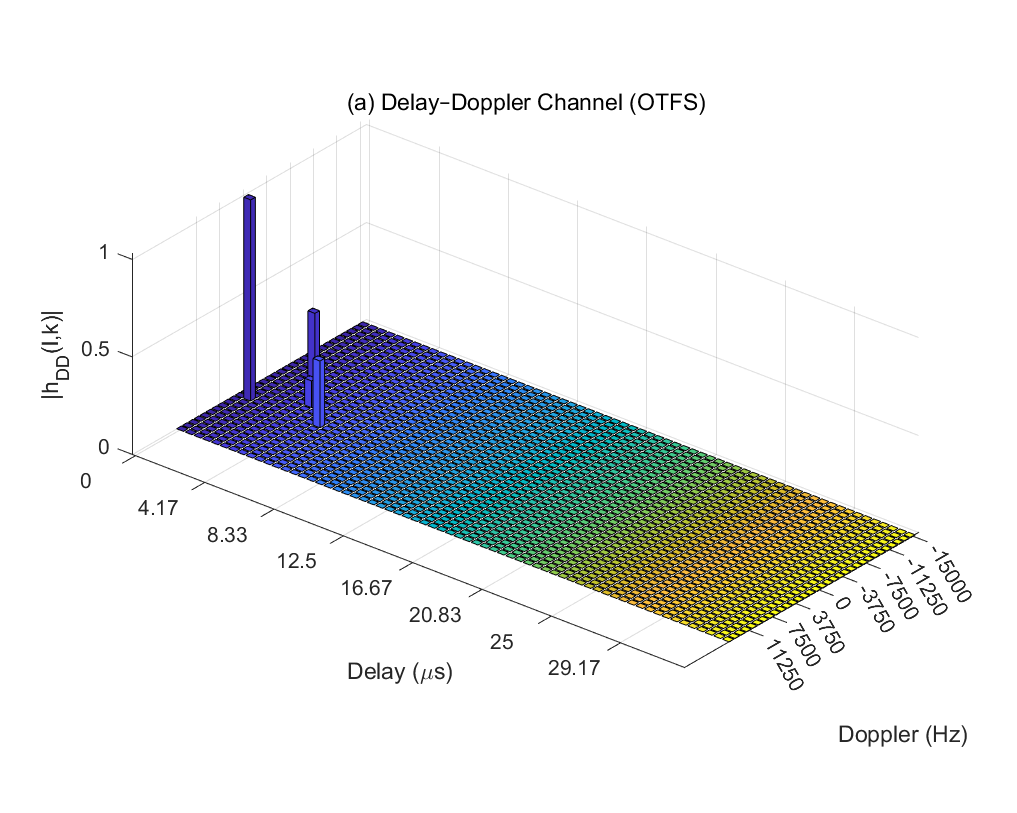}
        \caption{Delay-Doppler Channel}
        \label{fig:10a}
    \end{subfigure}
    \hfill
    \begin{subfigure}{0.45\textwidth}
        \centering
        \includegraphics[width=\textwidth]{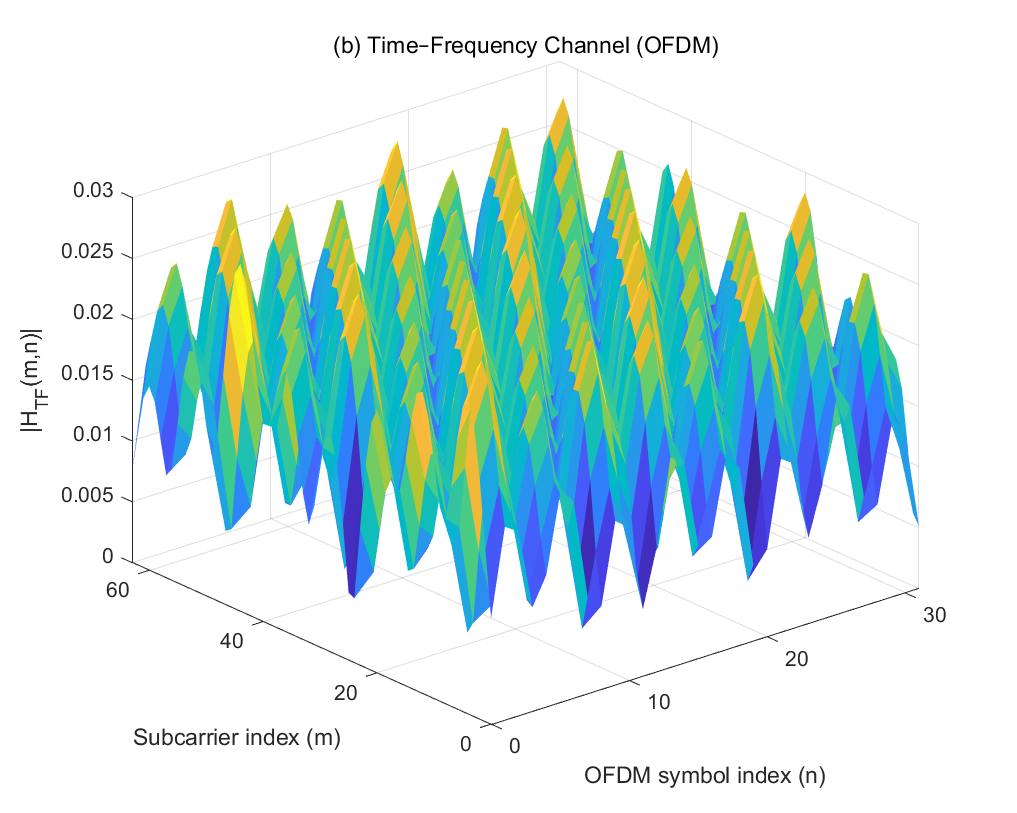}
        \caption{Time-Frequency Channel}%
        \label{fig:10b}
    \end{subfigure}
    \caption{Channel response in different domains}
    \label{fig:10}
\end{figure}

\textbf{Solution:} OTFS modulation mitigates these issues by operating in the Delay-Doppler domain\cite{ Wref71}, transforming time-varying channels into quasi-static ones. Moreover, RIS relies on accurate and timely CSI to configure its phase shifts effectively, making OTFS particularly suitable for scenarios where rapid and reliable channel estimation is critical. As shown in \ref{fig:10}, OTFS maintains well-defined impulse responses under high doppler, unlike OFDM's smeared time-frequency response.

Unless otherwise specified, the conceptual channel plots in Fig.~\ref{fig:10} are generated with
carrier frequency $f_c=20~\mathrm{GHz}$ and subcarrier spacing $\Delta f=30~\mathrm{kHz}$.
We consider an $M\times N=64\times 32$ OTFS/OFDM grid, yielding a frame duration
$T_{\mathrm{frame}}=N/\Delta f\approx 1.07~\mathrm{ms}$.
The excess delay spread is up to $5~\mu\mathrm{s}$, corresponding to a delay resolution
$1/(M\Delta f)\approx 0.52~\mu\mathrm{s}$.and Doppler resolution $\Delta f/N\approx 937.5$~Hz. A LEO satellite moves at $v_{\mathrm{sat}}\approx 7.6$~km/s and the UE speed is set to $v_{\mathrm{user}}=30$~m/s. While the raw doppler at Ka-band can be on the order of several hundred kHz, we assume that most of the deterministic component is pre-compensated using ephemeris, and Fig.~10 therefore focuses on the residual doppler spread within one frame. The cascaded channel is abstracted by $P=4$ dominant components, with excess delays $\tau_p\in[0.6,4.7]~\mu$s and residual dopplers $\nu_p\in[-5.0,5.0]$~kHz.
After bulk doppler pre-compensation, the residual doppler spread within a frame is set to
$\pm 5~\mathrm{kHz}$ to emulate high-mobility conditions.

\subsubsection{Seamless Handover}

The simultaneous mobility of satellites and users in LEO networks necessitates frequent and seamless handovers to maintain service continuity. This challenge is further exacerbated by the introduction of RIS, which increases the number of potential handover points, thereby complicating the decision-making process for selecting target links and determining optimal handover timing. These challenges require innovative solutions to ensure seamless connectivity and prevent service disruptions.

\textbf{Solution:} A general method is to use satellite ephemerids and current CSI to compute a look-ahead window of per-candidate beams and then trigger make-before-break when the projected links cross a hysteresis margin, enabling dual connectivity and soft combining. Hierarchical mobility management and software defined network (SDN)-based scheduling provide a scalable framework to decouple control and data, keeping RIS management on a reliable control link \cite{Wref65}\cite{Wref68}, thereby allowing flexible and coordinated handover strategies across heterogeneous links. To reduce signaling overhead, lightweight mechanisms employ asynchronous or event-driven operation \cite{Wref67}, which minimizes control message exchanges during rapid topology changes.

\subsubsection{Beam Tracking}

In LEO satellite networks, continuous beam realignment is required to maintain reliable links within communication sessions, even in the absence of handovers. The high mobility of both satellites and users causes rapid angular changes, which degrade signal quality and increase outage probability if not compensated. Closed-loop feedback mechanisms are limited by inherent propagation and processing latency, resulting in obsolete configuration updates before they are applied. 

\textbf{Solution:} Accurately forecasting user trajectories is challenging due to maneuver uncertainty, atmospheric disturbances, and measurement noise. The use of hybrid model-based and data-driven predictors enables the prediction of beam states and the scheduling of minimal probing. Physics-informed neural networks incorporate orbital dynamics and physical constraints to improve prediction accuracy in scenarios with relatively predictable movement patterns~\cite{Wref70}. For more dynamic or less predictable environments, generative AI based optimization techniques have been explored to synthesize likely trajectory patterns and proactively adjust beamforming strategies \cite{WrefAI}.

\textbf{Operational assumptions :}
We consider a nearly-passive RIS with phase-only control and a low-rate controller, i.e., the RIS does not perform high-rate baseband processing or per-element channel estimation. Instead, an effective cascaded CSI or an effective beam index is obtained via pilot probing and processed at the serving satellite payload or an associated gateway controller. To reduce dimensionality, the RIS can be operated with element grouping and a finite phase codebook. The receiver estimates the effective channel quality for a small set of candidate RIS codewords rather than attempting full per-element cascaded CSI.

\textbf{Example tracking loop and realistic update rates:}
Beam tracking can fuse (i) geometry-based prediction including satellite ephemeris,attitude and approximate UE location, (ii) communication-side measurements including timing, residual doppler, and best codeword index feedback, and when available (iii) auxiliary terminal sensors. A practical loop is: (1) predict the next beam direction and RIS codeword from geometry and the previous state; (2) probe a small set of $K$ candidate beams/codewords with short pilots; (3) measure and feed back the best index and link-quality metric; (4) update a lightweight filter e.g., Kalman-type filtering; (5) apply the selected satellite beam and RIS phases for the next interval. To bound signaling overhead, updates are event-driven: in steady-state route-aware LEO operation, differential updates on the order of a few times per second are typically sufficient, while short higher rate probing bursts are triggered only upon blockage, rapid prediction-error growth, or impending handover.

\subsection{Resource Allocation}

The inherent heterogeneity and scalability of satellite networks present challenges in ensuring quality of service (QoS) and optimizing resource allocation, primarily due to the disparity in resource availability \cite{bariah2022ris}. 
In RIS-assisted and satellite networks, resource allocation has been widely investigated, with a rich body of optimization-based, learning-based, and protocol-level solutions.

In the following, we identify how representative challenges manifest differently when RIS is integrated into satellite systems, and summarize corresponding solutions that are practically viable to provide implementation insight.

\subsubsection{Efficient Optimization}

Joint beamforming optimization and RIS configuration is a well-established topic in RIS-assisted networks. 

Considering the complex environment, transmission and reception conditions, and the presence of RIS elements, as well as various network resources, joint optimization of satellite beamforming and filtering vectors, RIS phase shifts, and other factors will be necessary \cite{ibrahim2023joint}.
These variables are tightly coupled and must satisfy power, QoS, and RIS unit-modulus constraints.
Specifically, each reflection coefficient must have a constant amplitude with tunable phases, making the problem non-convex and often NP-hard \cite{liu2021reconfigurable}. 
Moreover, multi-hop transmission in RIS-enhanced satellite networks further complicates the objective functions, involving more complex fractional signal-to-interference-plus-noise ratio (SINR) expressions and a computationally intensive decomposition process.

On the other hand, resource allocation can be shaped by where optimization is executed, i.e., onboard computation or ground computation. 
Specifically, long-horizon, high-dimensional optimizations, such as global beam scheduling \cite{xv2023joint}, RIS code-book generation \cite{kim2024joint}, constellation-wide spectrum reuse \cite{li2024game}, and QoS-aware resource slicing \cite{hugh2025resource}, are executed on the ground, where computational power is abundant and latency is less critical.
In contrast, fast-timescale tasks, including per-frame beam steering updates \cite{momani2025beam}, RIS element reconfiguration \cite{lin2022refracting}, link adaptation \cite{morel2015link}, and interference suppression decisions \cite{liu2025unintended}, must run onboard due to the tens to hundreds of milliseconds round-trip delay between satellites and ground stations. These latency constraints imply that onboard optimization must rely on lightweight inference models, pre-computed policies, or hierarchical decomposition rather than large centralized solvers.

Since LEO–ground round-trip times range from 30–80 ms, and cross-layer GEO–LEO exchanges may exceed 500 ms, closed-loop learning-based optimization cannot rely on instantaneous feedback. Reinforcement learning (RL) \cite{he2023hierarchical} and generative AI (GenAI)-based policy synthesis \cite{Wref70} therefore operate mainly in an offline or semi-offline manner on the ground, producing compact onboard-deployable policies, while only policy execution and minor corrections occur in real time on the satellite.

\textbf{Solution:} First, a hierarchical resource management architecture \cite{he2023hierarchical} can be incorporated. At the top level, a central controller performs coarse scheduling; at lower levels, local controllers handle real-time adjustments, reducing coordination latency and computation.
Moreover, the problem can be decomposed by alternating optimization \cite{li2024bistatic}, where the RIS phase shifts and amplitude coefficients are iteratively updated.
Penalty-based methods \cite{xu2025robust} or the successive convex approximation (SCA) \cite{jiang2025hybrid} can relax unit-modulus constraints.
Semi-definite relaxation (SDR) \cite{dai2024secrecy} approximates the RIS phase matrices, especially in static or quasi-static satellite scenarios.
Deep RL \cite{tang2022deep} can also replace exact optimization by following near-optimal schemes. 

While deep RL \cite{tang2022deep} and GenAI models \cite{he2023hierarchical,khoramnejad2025generative} offer strong approximation capability, their practical deployment can be constrained by (1) limited onboard memory and compute, (2) strict power budgets, (3) safety requirements that prevent exploratory actions in live networks, and (4) distribution shift, because satellite channels vary with orbit, geometry, and RIS placement.
As a result, current implementations emphasize offline training with policy distillation, where satellites only perform fast-inference using small-footprint models.

\subsubsection{Heterogeneous QoS}

As in conventional satellite and terrestrial systems, RIS-enhanced satellite networks must support diverse services,
including massive IoT, mobile broadband, ultra-reliable low-latency communications (URLLC), and vehicle/maritime links \cite{naaz2024empowering}.
Each use case imposes distinct QoS requirements, making resource management more complicated. Access characteristics also vary; ground IoT devices may require low-rate access \cite{ma2019high}, while aerial or maritime users demand high-throughput links\cite{salmeno2024reliable}. This imbalance between applications and performance demands in RIS-enhanced satellite networks makes it difficult to simultaneously optimize all QoS and access requirements.

\textbf{Solution:} A service-aware resource slicing framework \cite{ojaghi2022slicedran} can partition RIS-assisted resources for services depending on latency and reliability. For example, critical URLLC traffic can be assigned dedicated beams and robust RIS configurations with low-latency signaling paths. Priority-based scheduling \cite{xu2016priority} and utility-aware optimization \cite{das2021privacy} can dynamically reallocate resources. Services and users will be assigned specific orders to improve utilization efficiency and balance end-user performance.

\subsubsection{Network-Level Protocol}

In recent years, there are many existing SDN-based and NTN standardization research with protocol-level designs.
On this basis, rather than to redefine protocol architectures, RIS can be employed to introduce additional control endpoints and timescale separation.
To support efficient resource management in RIS-enhanced satellite networks, seamless coordination is required between satellites, RIS panels, and users. 
However, due to multi-hop characteristics brought about by RIS, conventional real-time control for satellites and BSs over RIS-assisted links must be reshaped.
Dynamic allocation demands frequent state exchange in cross-layer domains with disparate latency and bandwidth constraints \cite{zhao2024dynamic}. 
Moreover, protocol incompatibility between space and ground segments, that is, NTN and terrestrial networks, can hinder unified scheduling and QoS enforcement across different layers.

\textbf{Solution:} 
Lightweight RIS control signaling mechanisms can be integrated into satellite network protocols to mitigate overhead \cite{kunz2025lightweight}, operating asynchronously or event-triggered. 
Hierarchical control architectures based on an SDN \cite{bannour2017distributed} can abstract resources into programmable units for centralized or federated scheduling.

Furthermore, as real-time control of RIS and satellite beams must operate within the constraints of delayed command loops, protocol design prioritizes predictive configuration and event-triggered updates rather than continuous feedback.
By extending architectures/protocols such as the next-generation radio access network (NG-RAN) and the consultative committee for space data systems (CCSDS) \cite{de2020qos}, RIS signaling for feedback and updates can be compressed, latency-tolerant and standardized.

\text{}

It is worth mentioning that resource allocation in RIS-enhanced satellite networks should be viewed as an adaptation and re-layering of established methods, rather than a replacement. 
The key challenge is aligning optimization timescales, control overhead, and physical constraints imposed by satellite motion and active/passive RIS operations.

\subsection{Hardware Imperfections}

Introducing RIS into satellite networks brings hardware imperfections that are absent or less pronounced in conventional direct satellite links and, in some cases, amplify existing impairments due to the cascaded and high-dimensional nature of RIS-assisted channels. 
Across all deployment cases, practical RISs inevitably exhibit non-ideal reflection/transmission responses, including finite phase resolution, amplitude-phase coupling, inter-element mutual coupling,  radiation efficiency loss, and limited tuning range. 
These impairments reduce coherent combining gain, distort the intended beam pattern (e.g., elevated sidelobes), and may also bias channel estimation and beam training when the assumed RIS model deviates from hardware reality. 
In wideband systems, frequency-dependent distortions (e.g., beam squint and beam dispersion) further degrade array gain due to the typically frequency-flat phase control in many RIS architectures. 
Moreover, the need to configure a large number of elements creates non-trivial control overhead and reconfiguration latency, which can be particularly challenging under LEO dynamics.
Unlike terrestrial-only RIS deployments, spaceborne RISs (e.g., RIS as a satellite antenna or as an inter-satellite relay) additionally operate in a harsh space environment. Specifically, cumulative radiation and repeated sunlit/eclipse thermal cycling can induce slow parametric drifts (e.g., temperature- and dose-dependent variations in phase/amplitude response) and abrupt faults (e.g., transient upsets or latchups in control electronics), thereby degrading long-term beam-pattern fidelity unless compensated by calibration and fault-tolerant control. 
Table~\ref{tab:impairments_bands_cases} summarizes the dominant impairment sources across Ku/Ka, THz, and optical bands for the four deployment cases, highlighting which issues are mainly \emph{band-dependent} versus \emph{deployment-dependent}.


\begin{table*}[!t]
\centering
\caption{Checklist of dominant RIS hardware impairments across deployment cases and frequency bands, where $\checkmark$ means typically non-negligible, ``Opt." denotes optical-band RIS, and ``Satellite", ``Terminal", ``Inter-satellite", and ``Space-ground" denote RIS as a satellite antenna, a terminal antenna, an inter-satellite relay, and a space-ground relay, respectively.}
\label{tab:impairments_checklist}
\scriptsize
\setlength{\tabcolsep}{0.3pt}
\renewcommand{\arraystretch}{1.35}
\begin{tabular}{|>{\raggedright\arraybackslash}p{2.3cm}|c|c|c|c|c|c|c|c|c|c|c|c|}
\hline
& \multicolumn{3}{c|}{\textbf{Satellite}}
& \multicolumn{3}{c|}{\textbf{Terminal}}
& \multicolumn{3}{c|}{\textbf{Inter-satellite}}
& \multicolumn{3}{c|}{\textbf{Space-ground}} \\

& \textbf{Ku\!/\!Ka} & \textbf{THz} & \textbf{Opt.}
& \textbf{Ku\!/\!Ka} & \textbf{THz} & \textbf{Opt.}
& \textbf{Ku\!/\!Ka} & \textbf{THz} & \textbf{Opt.}
& \textbf{Ku\!/\!Ka} & \textbf{THz} & \textbf{Opt.} \\
\hline

\makecell[l]{Finite phase \\resolution}
& $\checkmark$ & $\checkmark$ & $\checkmark$ & $\checkmark$ & $\checkmark$ & $\checkmark$ &  & $\checkmark$ & $\checkmark$ & $\checkmark$ & $\checkmark$ &  $\checkmark$ \\
\hline
\makecell[l]{Amplitude-\\phase coupling}
& $\checkmark$ & $\checkmark$ & $\checkmark$ & $\checkmark$ & $\checkmark$ & $\checkmark$ &  & $\checkmark$ & $\checkmark$ & $\checkmark$ & $\checkmark$ &  $\checkmark$ \\
\hline
\makecell[l]{Mutual coupli-\\ng effect}
& $\checkmark$ & $\checkmark$ & $\checkmark$ & $\checkmark$ & $\checkmark$ & $\checkmark$ &  & $\checkmark$ & $\checkmark$ & $\checkmark$ & $\checkmark$ &  $\checkmark$ \\
\hline
\makecell[l]{Insertion and \\efficiency loss}
& $\checkmark$ & $\checkmark$ & $\checkmark$ & $\checkmark$ & $\checkmark$ & $\checkmark$ &  & $\checkmark$ & $\checkmark$ & $\checkmark$ & $\checkmark$ & $\checkmark$ \\
\hline
\makecell[l]{Tuning range\\limitation}
&   & $\checkmark$ & $\checkmark$ &   & $\checkmark$ & $\checkmark$ &   & $\checkmark$ & $\checkmark$ &   & $\checkmark$ & $\checkmark$ \\
\hline
\makecell[l]{Wideband beam\\squint or split}
&  $\checkmark$ & $\checkmark$ &  &   & $\checkmark$ &  &   & $\checkmark$ &  &   & $\checkmark$ &  \\
\hline
\makecell[l]{Fabrication-\\tolerance\\sensitivity}
&  & $\checkmark$ & $\checkmark$ &  & $\checkmark$ & $\checkmark$ &  & $\checkmark$ & $\checkmark$ &  & $\checkmark$ & $\checkmark$ \\
\hline
\makecell[l]{Packaging loss}
&  & $\checkmark$ & $\checkmark$ &  & $\checkmark$ & $\checkmark$ &  & $\checkmark$ & $\checkmark$ &  & $\checkmark$ & $\checkmark$ \\
\hline
\makecell[l]{Thermal drift\\and cycling}
& $\checkmark$ & $\checkmark$ & $\checkmark$ & $\checkmark$ & $\checkmark$ & $\checkmark$ &  & $\checkmark$ & $\checkmark$ & $\checkmark$ & $\checkmark$ & $\checkmark$ \\
\hline
\makecell[l]{Radiation effect\\(TID/SEE)}
& $\checkmark$ & $\checkmark$ & $\checkmark$ &  &  &  &  & $\checkmark$ & $\checkmark$ &  &  &  \\
\hline
\makecell[l]{Size and weight \\limitation}
& $\checkmark$ & $\checkmark$ & $\checkmark$ & $\checkmark$ & $\checkmark$ & $\checkmark$ &  & $\checkmark$ & $\checkmark$ &  &  &  \\
\hline
\makecell[l]{Vibration and \\jitter sensitivity}
&  & $\checkmark$ & $\checkmark$ &  & $\checkmark$ & $\checkmark$ &  & $\checkmark$ & $\checkmark$ &  & $\checkmark$ & $\checkmark$ \\
\hline
\makecell[l]{Blockage and\\pose variation}
&  &  &  & $\checkmark$ & $\checkmark$ & $\checkmark$ &  &  &  & $\checkmark$ & $\checkmark$ & $\checkmark$ \\
\hline
\makecell[l]{Contamination \\ and deposition}
&  &  & $\checkmark$ &  &  & $\checkmark$ &  &  & $\checkmark$ & $\checkmark$ & $\checkmark$ & $\checkmark$ \\
\hline
\makecell[l]{Reconfiguration\\latency}
& $\checkmark$ & $\checkmark$ & $\checkmark$ & $\checkmark$ & $\checkmark$ & $\checkmark$ &  & $\checkmark$ & $\checkmark$ & $\checkmark$ & $\checkmark$ & $\checkmark$ \\
\hline
\makecell[l]{Feedback and \\pilot overhead}
& $\checkmark$ & $\checkmark$ & $\checkmark$ & $\checkmark$ & $\checkmark$ & $\checkmark$ &  & $\checkmark$ & $\checkmark$ & $\checkmark$ & $\checkmark$ & $\checkmark$ \\
\hline
\makecell[l]{Eye-safety\\power limit}
&  &  &  &  &  & $\checkmark$ &  &  &  &  &  & $\checkmark$ \\
\hline
\end{tabular}
\end{table*}

\subsubsection{Phase Noise}

Practical RIS elements exhibit finite phase resolution, amplitude--phase coupling, element mismatch, and mutual coupling, which collectively manifest as phase perturbations (and possibly amplitude perturbations) relative to the ideal unit-modulus model. Such perturbations broaden the effective phase distribution and degrade coherent combining, thereby reducing array gain and potentially elevating undesired sidelobes. In RIS-assisted satellite links, these effects are further coupled with transceiver oscillator phase noise and Doppler-induced phase rotation, making robust modeling and compensation important, especially for narrow beams and high carrier frequencies.

\textbf{Solution:} A practical approach is to employ periodic calibration to track and compensate for the response of the RIS element. This can be achieved by (i) sparse sensing/telemetry (e.g., a small number of embedded detectors or occasional probing) to estimate beam-pattern deviations, and (ii) bias correction using look-up tables or lightweight online regression models. At the algorithmic level, robust beamforming designs that explicitly account for practical phase-error models (e.g., bounded errors or distributional models) can provide guaranteed performance via worst-case margins or probabilistic outage constraints.

\subsubsection{Wideband Effects}

In wideband RIS-enhanced satellite systems, beam squint and secondary beam dispersion become significant issues. Different frequency components of the same beam propagate in slightly different directions, leading to frequency-dependent misalignment, which is amplified in RIS-aided systems when the RIS phase profile is frequency-flat. The reflected/transmitted beams at different frequencies may deviate in both direction and power, reducing beamforming accuracy and array gain. This issue can be more pronounced in near-field or short-range RIS links (e.g., RIS-to-user relay segments), where spherical-wave effects and frequency-dependent focusing errors can further degrade gain.

\textbf{Solution:} Frequency-aware beamforming strategies (e.g., subband processing) can mitigate wideband distortions by optimizing beamforming per subband. RIS designs that incorporate quasi-true-time-delay (TTD) can more accurately approximate frequency-dependent phase profiles. In cases where full frequency-selective beamforming is impractical, combining RIS with movable antennas or adaptive satellite arrays can dynamically adjust the beam direction, minimizing misalignment. 

\subsubsection{Radiation and Thermal Effects}

For the two spaceborne deployment cases (RIS as a satellite antenna and RIS as an inter-satellite relay), the RIS surface, its bias network, and controller electronics must tolerate (i) total ionizing dose (TID) and (ii) single-event effects (SEE), in addition to (iii) repeated thermal cycling associated with frequent sunlit/eclipse transitions. In polar LEO, representative in-orbit dosimetry reports an order of $10^2$~Rad(Si)/year behind practical shielding thicknesses, implying that temperature- and dose-dependent response drift may accumulate over multi-year missions, while the design margin depends strongly on orbit and shielding. SEE can induce transient upsets or permanent failures in control and mixed-signal circuits, and the resulting configuration errors can directly translate into beam-pattern distortion. Thermal cycling and gradients can further perturb the element response through temperature-dependent material/electrical parameters and thermomechanical deformation of large-aperture panels, where even modest warping may cause noticeable phase-front distortions across the aperture.

\textbf{Solution:} From a wireless-system perspective, the key is to maintain beam-pattern fidelity under slow drift and occasional faults. This can be supported by: (i) radiation-aware control electronics (e.g., shielding, latchup protection, watchdog resets, redundancy, and basic error detection/correction in the RIS controller), (ii) temperature-aware compensation (e.g., embedding temperature sensors and using pre-characterized correction tables for phase/amplitude drifts), and (iii) periodic in-situ calibration using sparse probing and telemetry to refresh the effective RIS model used by beamforming and channel estimation. For ground-based RISs in the space-ground relay case, radiation is negligible, and the emphasis shifts to temperature/humidity stability and long-term outdoor aging; for terminal-mounted RISs, dominant thermal impairments often arise from device self-heating and rapid orientation changes rather than space radiation.

\subsubsection{Control Overhead}

RIS introduces a high-dimensional cascaded channel and extensive control signaling for massive elements, while its passive implementation offers limited onboard power and processing capability. Hence, pilot overhead and feedback can scale poorly if per-element states are tracked. Moreover, reconfiguration latency is non-negligible: finite-rate control links and switching times (potentially on the order of milliseconds or longer, depending on the tuning technology) can limit the update rate relative to LEO dynamics, thereby impacting beam tracking and handover.

\textbf{Solution:} A low bit rate but reliable control link with robust modulation is needed, e.g., an ultra-high frequency (UHF) command link to a HAP-based RIS or a wired fiber link to a building RIS. Then, estimate the effective cascaded channels using compressed pilots and element-grouping strategies. Designing the control protocol and event-driven scheduling (e.g., blockage/handover triggers) for differential updates to keep control below a few updates per second, which is adequate for route-aware LEO operation, the system can adopt semi-static codebooks indexed by satellite ephemeris, user sector, and platform pose, partially updating rather than the full state, and sector-based configuration rather than individual elements. 

\section{Future Directions and Discussion}

\subsection{GenAI for RIS-enhanced Satellite Networks}

This subsection explores the potential role of generative models as auxiliary tools in satellite network control and optimization, rather than as standalone or fully autonomous decision-makers.
GenAI is expected to offer a transformative paradigm shift from deterministic control to data-driven adaptability for RIS-enhanced satellite networks. Considering the high-dimensional, dynamic, and partially observable characteristics of satellite networks, GenAI models, such as generative diffusion models, large language models (LLMs), and vision transformers, can synthesize representative channel environments, generate adaptive beamforming strategies, and predict user or traffic behavior under uncertainty \cite{zhang2024generative}.

To ensure real-time and scalable deployment of RIS-enhanced satellite networks, lightweight generative architectures must be designed for onboard inference within satellite payloads or RIS controllers, which may also involve federated or split learning paradigms. These architectures can reduce reliance on ground-based control, provide greater design and application flexibility, and support local policy updates and low-latency decisions.
Moreover, due to the dynamic nature of satellite networks, generative models can be trained on historical satellite telemetry, orbital dynamics, and RIS control traces. In this sense, multi-dimensional factors, including waveform adaptation, RIS topology reconfiguration, and mobility-aware resource allocation, can be jointly designed from a system-level perspective, thereby achieving higher robustness, spectral efficiency, and other benefits beyond conventional heuristics.
In the future, GenAI can be anticipated to serve not only as a tool for optimization \cite{khoramnejad2025generative}, but also as a foundation to enable autonomous and cognitive RIS-enhanced satellite networks.

\subsubsection{Practical Limitations}
However, generative models face constraints in satellite systems.
First, training-data scarcity is inherent since satellite–RIS–user geometries vary across orbits, environments, and hardware platforms, causing limited representativeness of historical datasets and distribution shift.
Second, generative diffusion models, transformers, and LLM-based controllers require orders of magnitude more computation and memory than classical solvers, while the onboard power, thermal, and processor constraints are strict.
Third, closed-loop validation is difficult because unsafe generative outputs induce suboptimal beamforming, pointing errors, or link outages, and satellites cannot easily support exploratory online learning.

These factors suggest that GenAI can currently be most suitable for offline or semi-offline policy generation, where only lightweight inference models are deployed onboard.

\subsubsection{Use Case: Generative Channel \& Mobility for Optimization}
As an illustrative example, we consider employing generative models to synthesize satellite–RIS–user channel realizations or user mobility traces that augment scarce real measurements. In a standard optimization loop, this operates as follows:
\begin{enumerate}
    \item Generative diffusion model trained offline on historical telemetry, orbital dynamics, and blockage maps;
    \item Model produces diverse but physically consistent synthetic samples;
    \item Samples fed into conventional optimization methods, e.g., alternating optimization, SCA, SDR, or RL-based resource allocation, to improve robustness;
    \item Resulting optimized policies compressed or distilled into a lightweight inference model that can run onboard.
\end{enumerate}

In this sense, GenAI supports but does not replace classical optimization. This will provide a practical way to improve resilience under data scarcity and variable network geometry.


\subsection{Multifunctional RIS Architectures for Satellite Networks}
A simultaneously transmitting and reflecting RIS divides the incident energy into programmable reflected and refracted parts, beaming feeder power down to Earth while forwarding part of it to an ISL, or sitting on a facade to serve both outdoor and indoor users simultaneously. Active and hybrid RISs embed low-noise amplifiers behind every element, or selected elements, turning the RIS into a low-complexity relay that provides improved link gain, which is critical in the Ka/THz bands. Conformal and lens RISs wrapped around curved structures enable near-omnidirectional electronic beam steering. Ground terminals already utilize millimeter-thin steerable metasurface lenses illuminated by a single feed, delivering dish-like gain with no moving parts; the same concept can be applied to a satellite for panoramic coverage. Flexible RIS with printable plastic or textile metasurfaces enables balloons, roofs, or UAV wings to act as steerable reflectors, creating ultralight, easily deployable apertures for emergency or backhaul links. High-frequency RISs, including nanostructured THz and optical metasurfaces, promise sub-milliradian steering without the need for gimbals. An optical RIS in LEO could instantaneously repoint laser links; an active THz surface integrated with solar panels could offset severe path loss while correcting attitude jitter. Holographic RISs with subwavelength spacing provide continuous-aperture control and near-linear capacity scaling with area, suggesting the potential for ultra-high-throughput gateways without the need for thousands of RF chains.

To make this landscape more actionable, it is helpful to attach representative physical scales, performance expectations, and a readiness view. At Ku/Ka, most electronically reconfigurable reflectarray/transmitarray-type metasurfaces employ an element pitch on the order of $\lambda/2$ with a millimeter scale, while some transmissive/lens-type surfaces adopt sub-wavelength sampling to improve wavefront control. For these aperture-type implementations, the realized gain is primarily set by the electrical area and can be approximated by $G \approx \eta 4\pi A/\lambda^2$, where $\eta$ captures radiation efficiency; thus, moving from Ku to Ka rapidly boosts directivity for a fixed form factor, while wide-angle scanning and wideband operation are typically limited by element dispersion and scan loss. In practice, Ku/Ka metasurface apertures can support multi-GHz bandwidths ($\sim$10\% fractional), while multifunctional variants (e.g., shared-aperture transmit/reflect) generally incur additional splitting/duplexing and isolation penalties, which can translate into several dB of effective link-budget loss unless carefully engineered. From a feasibility standpoint, near-term deployable candidates include passive Ku/Ka-band reflective or transmissive metasurfaces and their conformal or flexible variants, provided that calibration is used to compensate for deformation and thermal drift; active and hybrid surfaces are more plausibly mid-term contingent options for Ka/mmWave when link margins are tight, but their net benefit depends on noise figure, stability, power delivery, and thermal or radiation hardening. By contrast, sub-THz, THz, and optical RIS-like concepts remain longer-term exploratory: at THz, insertion loss and phase dispersion, packaging/bias routing, phase noise, synchronization sensitivity, and thermal/power limits dominate; at optical frequencies, micron-scale pixel or element pitch leads to challenging driver scaling, and dispersion or achromaticity, power handling, radiation and thermal stability, and pointing and jitter stabilization are the key obstacles for scaling reconfigurable wavefront control from millimeter-scale devices to space-grade, large-aperture beam-steering payloads.

\subsection{RIS-Enhanced Satellite ISAC}
The space-air-ground-sea integrated network enables multi-domain collaboration, wide coverage, and heterogeneous access for ISAC. However, traditional ground-based radar sensing systems suffer from coverage limitations and performance degradation under complex propagation and fading. RIS-enhanced satellite networks mitigate blind spots, as depicted in the previous sections, and enable seamless ISAC coverage in obstructed environments. Through dynamic reflection control, RIS forms additional communication and sensing paths \cite{chepuri2023integrated}, reconfigures the environment, extends coverage, and supports multi-perspective perception in cooperation with satellite systems. In this way, satellite-enabled ISAC can adapt to time-varying channels, improving the accuracy of real-time sensing \cite{ge2024ris}. RISs mounted in buildings can also direct satellite signals indoors while reflecting sensing beams toward urban targets \cite{wang2023stars}. In environments where global navigation satellite system (GNSS) is denied, as LEO satellites may still encounter blockage, multipath, and doppler effects \cite{dureppagari2023ntn}, RIS improves propagation and provides high-resolution geometric measurements, supporting advanced localization models for seamless indoor–outdoor sensing coverage \cite{wang2024beamforming,saleh20246g,zheng2023leo}. \textcolor{orange}{Existing works on RIS-aided LEO positioning indicate that meter-level localization accuracy is achievable under favorable SNR and geometry, e.g., RIS-aided single-LEO localization reports meter-level positioning error at 30 dB SNR\cite{saleh20246g}.} \textcolor{orange}{It is worth noting that RIS-enhanced satellite ISAC involves sensing–communication trade-offs: RIS configurations that minimize position error bound may not maximize communication throughput, especially under specific constraints such as high doppler, limited coherence time and satellite mobility. }

\subsection{RIS-Enhanced Satellite IoT}
In satellite IoT networks, efficiently supporting and managing massive access is challenging due to the large number of terminals and limited access resources, and the number of ground devices often exceeds the capacity of the satellites. RIS can reduce the coverage burden for satellite networks, allowing flexible adaptation of coverage to the spatial distribution of IoT devices \cite{tekbiyik2021energy}. Deploying RIS near terminals enables controllable, directionally distinct reflection paths, enhancing spatial separability and angular discrimination in dense device scenarios. However, improving the coverage of the physical layer alone is insufficient. Joint RIS design, combined with advanced access schemes, can fully leverage spatial control to support multiple simultaneous accesses in a limited spectrum \cite{zou2023machine}. For random, low-data short-term activations, integrating RIS with grant-free access reduces collision probability and delay \cite{marinello2025grant}, indicating a strong potential for satellite IoT under strict resource constraints. Energy efficiency is also critical due to limited terminal battery life and high satellite power costs. The passive nature of RIS reduces hardware costs, eliminates complex satellite processing, and extends the battery life of IoT devices. \textcolor{orange}{RIS-assisted propagation enhancement enables one- to two-order-of-magnitude reductions in IoT transmit power under fixed QoS constraints \cite{tekbiyik2021energy}, while system-level energy savings on the order of tens of percent have been reported even in more complex satellite–UAV–IoT networks \cite{lukito2024integrated}. Therefore, RIS was positioned as a key enabler for sustainable and energy-efficient satellite IoT systems.}

\section{Conclusion}

RISs are compelling in satellite networks not as a universal complement for denser constellations or larger active arrays, but as a way to convert favorable geometry into a controllable network resource. Across the four deployment archetypes, a consistent physical lesson emerges: the achievable gain is set by aperture, efficiency, and two-hop loss, which in turn determines whether an RIS behaves like a lightweight high-gain aperture for satellite- or terminal-mounted antenna, a short-range link reflector for inter-satellite relay, or a localized coverage shaper that creates virtual-LoS propagation for space–ground relay. This viewpoint turns deployment from an implementation detail into a prerequisite design choice.

At the system level, RIS-enabled satellite networking can be understood in terms of two unifying functions. First, RISs restore connectivity by constructing virtual LoS paths that maintain continuity across blockage-dominated outdoor-to-indoor transitions and along mobility trajectories. Second, RISs provide angular selectivity that complements scheduling and active beamforming by mitigating interference in space, thereby broadening feasible spectrum-sharing regimes within a layer, across layers, and across satellite–terrestrial coexistence. Framing RIS benefits through these two functions helps distinguish cases in which RISs provide new leverage from those in which they primarily achieve conventional beamforming.

Realizing these gains at scale demands co-design across three layers: (i) placement and motion planning that respects visibility windows and mechanical constraints; (ii) control and optimization that explicitly budgets overhead and latency, favoring predictive, event-driven updates over continuous feedback; and (iii) hardware-aware operation that accounts for wideband effects and long-term drift or faults under radiation and thermal cycling. Looking ahead, progress will likely be driven by field demonstrations and control-native approaches, ranging from multifunctional RIS hardware to learning- and GenAI-assisted orchestration, that jointly close the loop among sensing, configuration, and network objectives in dynamic non-terrestrial environments.

\bibliography{references}

@article{Khennoufa2025MultiLayerNTN,
  author  = {Faical Khennoufa and Khelil Abdellatif and Halim Yanikomeroglu and Metin Ozturk and Taissir Elganimi and Ferdi Kara},
  title   = {A Multi-Layer Non-Terrestrial Networks Architecture for 6G and Beyond Under Realistic Conditions and With Practical Limitations},
  journal = {IEEE Internet of Things Magazine},
  year    = {2025},
  volume  = {8},
  number  = {5},
  pages   = {136--143},
  month   = {sep},
  doi     = {10.1109/MIOT.2025.3575923}
}

@article{Khan2025BeyondDiagonalRISNTN,
  author  = {W. U. Khan and A. Mahmood and M. A. Jamshed and E. Lagunas and M. Ahmed and S. Chatzinotas},
  title   = {Beyond Diagonal RIS for 6G Non-Terrestrial Networks: Potentials and Challenges},
  journal = {IEEE Network},
  year    = {2025},
  volume  = {39},
  number  = {1},
  pages   = {80--89},
  month   = {jan},
  doi     = {10.1109/MNET.2024.3480332}
}

@article{Cao2021ConvergedRISMEC,
  author  = {Xuelin Cao and Bo Yang and Chongwen Huang and Chau Yuen and Yan Zhang and Dusit Niyato and Zhu Han},
  title   = {Converged Reconfigurable Intelligent Surface and Mobile Edge Computing for Space Information Networks},
  journal = {IEEE Network},
  year    = {2021},
  volume  = {35},
  number  = {4},
  pages   = {42--48},
  month   = {jul},
  doi     = {10.1109/MNET.011.2100049}
}

@article{Liu2024Empowering6GNTNIoT,
  author  = {Q. Liu and Z. Feng and D. Chen and F. Tan and C. He},
  title   = {Empowering 6G Non-Terrestrial Networks With Intelligent Reflection Technologies for IoT Applications},
  journal = {IEEE Network},
  year    = {2024},
  volume  = {38},
  number  = {4},
  pages   = {96--102},
  month   = {jul},
  doi     = {10.1109/MNET.2024.3381580}
}

@article{Xu2021IRSSAGIN,
  author  = {S. Xu and J. Liu and T. K. Rodrigues and N. Kato},
  title   = {Envisioning Intelligent Reflecting Surface Empowered Space-Air-Ground Integrated Network},
  journal = {IEEE Network},
  year    = {2021},
  volume  = {35},
  number  = {6},
  pages   = {225--232},
  month   = {nov},
  doi     = {10.1109/MNET.011.2100007}
}

@article{Wu2025FLSTARRISSAGIN,
  author  = {M. Wu and others},
  title   = {Federated Learning in STAR-RIS Aided SAGINs},
  journal = {IEEE Communications Standards Magazine},
  year    = {2025},
  volume  = {9},
  number  = {2},
  pages   = {23--29},
  month   = {jun},
  doi     = {10.1109/MCOMSTD.2025.3569008}
}

@article{Saleh2025TNNTNLocalization,
  author  = {S. Saleh and others},
  title   = {Integrated 6G TN and NTN Localization: Challenges, Opportunities, and Advancements},
  journal = {IEEE Communications Standards Magazine},
  year    = {2025},
  volume  = {9},
  number  = {2},
  pages   = {63--71},
  month   = {jun},
  doi     = {10.1109/MCOMSTD.2025.3569014}
}

@article{Kaushik2024ISACIoT,
  author  = {A. Kaushik and others},
  title   = {Integrated Sensing and Communications for IoT: Synergies with Key 6G Technology Enablers},
  journal = {IEEE Internet of Things Magazine},
  year    = {2024},
  volume  = {7},
  number  = {5},
  pages   = {136--143},
  month   = {sep},
  doi     = {10.1109/IOTM.001.2400052}
}

@article{Xiao2024LEOSAN,
  author  = {Z. Xiao and others},
  title   = {LEO Satellite Access Network (LEO-SAN) Toward 6G: Challenges and Approaches},
  journal = {IEEE Wireless Communications},
  year    = {2024},
  volume  = {31},
  number  = {2},
  pages   = {89--96},
  month   = {apr},
  doi     = {10.1109/MWC.011.2200310}
}

@article{Luo2024LEOVLEO6G,
  author  = {X. Luo and H.-H. Chen and Q. Guo},
  title   = {LEO/VLEO Satellite Communications in 6G and Beyond Networks--Technologies, Applications, and Challenges},
  journal = {IEEE Network},
  year    = {2024},
  volume  = {38},
  number  = {5},
  pages   = {273--285},
  month   = {sep},
  doi     = {10.1109/MNET.2024.3353806}
}

@article{Ye2022NonterrestrialRISSProcIEEE,
  author  = {J. Ye and J. Qiao and A. Kammoun and M.-S. Alouini},
  title   = {Nonterrestrial Communications Assisted by Reconfigurable Intelligent Surfaces},
  journal = {Proceedings of the IEEE},
  year    = {2022},
  volume  = {110},
  number  = {9},
  pages   = {1423--1465},
  month   = {sep},
  doi     = {10.1109/JPROC.2022.3169690}
}

@article{Wu2024DRLRISIoTMag,
  author  = {M. Wu and K. Guo and X. Li and A. Nauman and K. An and J. Wang},
  title   = {Optimization Design in RIS-Assisted Integrated Satellite-UAV-Served 6G IoT: A Deep Reinforcement Learning Approach},
  journal = {IEEE Internet of Things Magazine},
  year    = {2024},
  volume  = {7},
  number  = {1},
  pages   = {12--18},
  month   = {jan},
  doi     = {10.1109/IOTM.001.2300111}
}

@article{Umer2025RISAerialNTNDRL,
  author  = {M. Umer and M. A. Mohsin and A. Kaushik and Q.-U.-A. Nadeem and A. A. Nasir and S. A. Hassan},
  title   = {Reconfigurable Intelligent Surface-Assisted Aerial Nonterrestrial Networks: An Intelligent Synergy With Deep Reinforcement Learning},
  journal = {IEEE Vehicular Technology Magazine},
  year    = {2025},
  volume  = {20},
  number  = {1},
  pages   = {55--64},
  month   = {mar},
  doi     = {10.1109/MVT.2024.3524745}
}

@article{Khan2024RIS6GNTNIoTMag,
  author  = {W. U. Khan and A. Mahmood and C. K. Sheemar and E. Lagunas and S. Chatzinotas and B. Ottersten},
  title   = {Reconfigurable Intelligent Surfaces for 6G Non-Terrestrial Networks: Assisting Connectivity from the Sky},
  journal = {IEEE Internet of Things Magazine},
  year    = {2024},
  volume  = {7},
  number  = {1},
  pages   = {34--39},
  month   = {jan},
  doi     = {10.1109/IOTM.001.2300208}
}

@article{Tekbiyik2022RISActionNTN,
  author  = {K. Tekb{\i}y{\i}k and G. K. Kurt and A. R. Ekti and H. Yanikomeroglu},
  title   = {Reconfigurable Intelligent Surfaces in Action for Nonterrestrial Networks},
  journal = {IEEE Vehicular Technology Magazine},
  year    = {2022},
  volume  = {17},
  number  = {3},
  pages   = {45--53},
  month   = {sep},
  doi     = {10.1109/MVT.2022.3168995}
}

@article{Bariah2023RISSAGINIEEE_Network,
  author  = {Lina Bariah and Lina Mohjazi and Hanaa Abumarshoud and Bassant Selim and Sami Muhaidat and Mallik Tatipamula and Muhammad Ali Imran and Harald Haas},
  title   = {RIS-Assisted Space-Air-Ground Integrated Networks: New Horizons for Flexible Access and Connectivity},
  journal = {IEEE Network},
  year    = {2023},
  volume  = {37},
  number  = {3},
  pages   = {118--125},
  month   = {may},
  doi     = {10.1109/MNET.123.2100761}
}

@article{Toka2024RISEmpoweredLEOComMag,
  author  = {M. Toka and E. Lagunas and M. A. Jamshed and S. Chatzinotas and B. Ottersten},
  title   = {RIS-Empowered LEO Satellite Networks for 6G: Promising Usage Scenarios and Future Directions},
  journal = {IEEE Communications Magazine},
  year    = {2024},
  volume  = {62},
  number  = {11},
  pages   = {128--135},
  month   = {nov},
  doi     = {10.1109/MCOM.002.2300554}
}

@article{Liu2024COMST_RIS_SATN_PDNOMA,
  author  = {R. Liu and others},
  title   = {RIS-Empowered Satellite-Aerial-Terrestrial Networks With PD-NOMA},
  journal = {IEEE Communications Surveys \& Tutorials},
  year    = {2024},
  volume  = {26},
  number  = {4},
  pages   = {2258--2289},
  doi     = {10.1109/COMST.2024.3393612}
}

@article{Khoshafa2025SecuredSAGIN,
  author  = {M. H. Khoshafa and F. Bueno and T. M. N. Ngatched and M. Di Renzo},
  title   = {RIS-Empowered Secured Space-Air-Ground Integrated Networks: Opportunities and Challenges},
  journal = {IEEE Communications Magazine},
  year    = {2025},
  volume  = {63},
  number  = {6},
  pages   = {130--136},
  month   = {jun},
  doi     = {10.1109/MCOM.001.2400398}
}

@article{Jamshed2024AirborneNTNRISIoT,
  author  = {M. A. Jamshed and others},
  title   = {Synergizing Airborne Non-Terrestrial Networks and Reconfigurable Intelligent Surfaces-Aided 6G IoT},
  journal = {IEEE Internet of Things Magazine},
  year    = {2024},
  volume  = {7},
  number  = {2},
  pages   = {46--52},
  month   = {mar},
  doi     = {10.1109/IOTM.001.2300242}
}

@article{Trinh2025QuantumSAGINOpticalRIS,
  author  = {P. V. Trinh and S. Sugiura and C. Xu and L. Hanzo},
  title   = {Toward Quantum SAGINs Harnessing Optical RISs: Applications, Advances, and the Road Ahead},
  journal = {IEEE Network},
  year    = {2025},
  volume  = {39},
  number  = {3},
  pages   = {215--222},
  month   = {may},
  doi     = {10.1109/MNET.2025.3536848}
}

@article{Ramezani2023RISEnhancedINTENT,
  author  = {P. Ramezani and B. Lyu and A. Jamalipour},
  title   = {Toward RIS-Enhanced Integrated Terrestrial/Non-Terrestrial Connectivity in 6G},
  journal = {IEEE Network},
  year    = {2023},
  volume  = {37},
  number  = {3},
  pages   = {178--185},
  month   = {may},
  doi     = {10.1109/MNET.116.2200060}
}

@article{Wu2025URLLC_DRL_RIS_SAGIN,
  author  = {M. Wu and K. Guo and X. Li and M. Asif and C. Han and K. M. Rabie},
  title   = {URLLC for DRL-RIS-AIDED SAGINs: New Mentalities, Trends and Preliminary Solutions},
  journal = {IEEE Communications Standards Magazine},
  year    = {2025},
  volume  = {9},
  number  = {1},
  pages   = {6--12},
  month   = {mar},
  doi     = {10.1109/MCOMSTD.0001.2400004}
}

@article{tekbiyik2022reconfigurableM,
  title={Reconfigurable {I}ntelligent {S}urfaces in {A}ction for {N}onterrestrial {N}etworks},
  author={Tekb{\i}y{\i}k, K{\"u}r{\c{s}}at and Kurt, G{\"u}ne{\c{s}} Karabulut and Ekti, Ali Riza and others},
  journal={IEEE Veh. Technol. Mag.},
  volume={17},
  number={3},
  pages={45--53},
  year={2022},
  publisher={IEEE}
}

@article{amodu2024technical,
  title={Technical {A}dvancements {T}owards {RIS}-{A}ssisted {NTN}-{B}ased {TH}z {C}ommunication for {6G} and {B}eyond},
  author={Amodu, Oluwatosin Ahmed and Nordin, Rosdiadee and Abdullah, Nor Fadzilah and others},
  journal={IEEE Access},
  volume={12},
  pages={183153-183181},
  year={2024},
  publisher={IEEE}
}

@article{shang2025channel,
  title={Channel {M}odeling and {R}ate {A}nalysis of {O}ptical {I}nter-{S}atellite {L}ink ({OISL})}, 
  author={Shang, Bodong and Zhang, Shuo and Wong, Zi Jing},
  journal={IEEE Trans. Veh. Technol.}, 
  year={2025},
  month={Mar.},
  volume={},
  number={},
  pages={1-6},
}

@ARTICLE{li2025advancing,
  author={Li, Xiangyu and Shang, Bodong},
  journal={IEEE Netw.}, 
  title={{A}dvancing {M}ulti-{C}onnectivity in {S}atellite-{T}errestrial {I}ntegrated {N}etworks: {A}rchitectures, {C}hallenges, and {A}pplications}, 
  year={2025},
  volume={},
  number={},
  pages={1-1},
  month={Feb.},
  publisher={IEEE}
}

@ARTICLE{shang2024multi,
  author={Shang, Bodong and Li, Xiangyu and Li, Zhuhang and others},
  journal={IEEE Open J. Commun. Soc.}, 
  title={{M}ulti-{C}onnectivity {B}etween {T}errestrial and {N}on-{T}errestrial {MIMO} {S}ystems}, 
  year={2024},
  volume={5},
  number={},
  pages={3245-3262},
  month={May},
  publisher={IEEE}
}

@inproceedings{shang2025spectrum_c,
  title={{S}pectrum {S}haring in {6G} {S}pace-{G}round {I}ntegrated {N}etworks: {A} {G}round {P}rotection {Z}one-{B}ased {D}esign},
  author={Shang, Bodong and Li, Xiangyu and Wang, Zheng and others},
  booktitle={Proc. Int. Conf. Intell. Commun. Netw. (ICN)},
  pages={40--45},
  year={2025},
  month={Feb.},
  organization={IEEE}
}

@article{shang2025spectrum_m,
  title={{S}pectrum {S}haring in {S}atellite-{T}errestrial {I}ntegrated {N}etworks: {F}rameworks, {A}pproaches, and {O}pportunities},
  author={Shang, Bodong and Wang, Zheng and Li, Xiangyu and others},
  journal={IEEE Netw.},
  year={2025},
  volume={},
  number={},
  pages={1-1},
  month={Sep.},
  publisher={IEEE}
}

@article{kunz2025lightweight,
  title={{L}ightweight {S}ecurity for {A}mbient-{P}owered {P}rogrammable {R}eflections with {R}econfigurable {I}ntelligent {S}urfaces},
  author={Kunz, Andreas and Baskaran, Sheeba Backia Mary and Alexandropoulos, George C},
  journal={arXiv:2501.09005. [Online]. Available: https://arxiv.org/abs/2501.09005},
  year={2025}
}

@article{kim2024spectrum,
  title={Spectrum {S}haring {B}etween {L}ow {E}arth {O}rbit {S}atellite and {T}errestrial {N}etworks: {A} {S}tochastic {G}eometry {P}erspective {A}nalysis},
  author={Kim, Daeun and Park, Jeonghun and Choi, Jinseok and others},
  journal={arXiv:2408.12145. [Online]. Available: https://arxiv.org/abs/2408.12145},
  year={2024}
}

@article{bariah2022ris,
  title={RIS-{A}ssisted {S}pace-{A}ir-{G}round {I}ntegrated {N}etworks: {N}ew {H}orizons for {F}lexible {A}ccess and {C}onnectivity},
  author={Bariah, Lina and Mohjazi, Lina and Abumarshoud, Hanaa and others},
  journal={IEEE Netw.},
  volume={37},
  number={3},
  pages={118--125},
  year={2022},
  month={Sep.},
  publisher={IEEE}
}

@article{li2024game,
  title={A {G}ame {T}heory-{B}ased {D}istributed {D}ownlink {S}pectrum {S}haring {M}ethod in {L}arge-{S}cale {H}ybrid {S}atellite {C}onstellations},
  author={Li, Wei and Jia, Luliang and Chen, Quan and others},
  journal={IEEE Trans. Commun.},
  year={2024},
  month={Aug.},
  volume={72},
  number={8},
  publisher={IEEE}
}

@article{he2023hierarchical,
  title={Hierarchical {C}ross-{D}omain {S}atellite {R}esource {M}anagement: {A}n {I}ntelligent {C}ollaboration {P}erspective},
  author={He, Hongmei and Zhou, Di and Sheng, Min and others},
  journal={IEEE Trans. Commun.},
  volume={71},
  number={4},
  pages={2201--2215},
  year={2023},
  month={Apr.},
  publisher={IEEE}
}

@article{bannour2017distributed,
  title={{D}istributed {SDN} {C}ontrol: {S}urvey, {T}axonomy, and {C}hallenges},
  author={Bannour, Fetia and Souihi, Sami and Mellouk, Abdelhamid},
  journal={IEEE Commun. Surveys Tuts.},
  volume={20},
  number={1},
  pages={333--354},
  year={2017},
  month={1st Quart.,},
  publisher={IEEE}
}

@article{de2020qos,
  title={{QoS} {O}ptimisation of {eMBB} {S}ervices in {C}onverged {5G}-{S}atellite {N}etworks},
  author={De Cola, Tomaso and Bisio, Igor},
  journal={IEEE Trans. Veh. Technol.},
  volume={69},
  number={10},
  pages={12098--12110},
  year={2020},
  month={Oct.},
  publisher={IEEE}
}

@article{zhao2024dynamic,
  title={Dynamic {R}esource {A}llocation for {M}ulti-{S}atellite {C}ooperation {N}etworks: {A} {D}ecentralized {S}cheme {U}nder {S}tatistical {CSI}},
  author={Zhao, Meihui and Yu, Hanxiao and Pan, Jianxiong and others},
  journal={IEEE Access},
  volume={12},
  pages={15419--15437},
  year={2024},
  month={Jan.},
  publisher={IEEE}
}

@article{xu2016priority,
  title={{P}riority-{B}ased {C}onstructive {A}lgorithms for {S}cheduling {A}gile {E}arth {O}bservation {S}atellites with {T}otal {P}riority {M}aximization},
  author={Xu, Rui and Chen, Huaping and Liang, Xinle and Wang, Huimin},
  journal={Expert Syst. Appl.},
  volume={51},
  pages={195--206},
  year={2016},
  publisher={Elsevier}
}

@article{das2021privacy,
  title={Privacy and {U}tility {A}ware {D}ata {S}haring for {S}pace {S}ituational {A}wareness {F}rom {E}nsemble and {U}nscented {K}alman {F}iltering {P}erspective},
  author={Das, Niladri and Bhattacharya, Raktim},
  journal={IEEE Trans. Aerosp. Electron. Syst.},
  volume={57},
  number={2},
  pages={1162--1176},
  year={2021},
  month={Apr.},
  publisher={IEEE}
}

@article{ojaghi2022slicedran,
  title={{S}liced{RAN}: {S}ervice-{A}ware {N}etwork {S}licing {F}ramework for {5G} {R}adio {A}ccess {N}etworks},
  author={Ojaghi, Behnam and Adelantado, Ferran and Antonopoulos, Angelos and Verikoukis, Christos},
  journal={IEEE Syst. J.},
  volume={16},
  number={2},
  pages={2556--2567},
  year={2022},
  month={Jun.},
  publisher={IEEE}
}

@article{naaz2024empowering,
  title={{E}mpowering the {V}ehicular {N}etwork with {RIS} {T}echnology: {A} {S}tate-of-the-{A}rt {R}eview},
  author={Naaz, Farheen and Nauman, Ali and Khurshaid, Tahir and Kim, Sung-Won},
  journal={Sensors},
  volume={24},
  number={2},
  pages={337},
  year={2024},
  month={Jan.},
  publisher={MDPI}
}

@article{ma2019high,
  title={{High-{R}eliability and {L}ow-latency Wireless Communication for Internet of Things: Challenges, Fundamentals, and Enabling Technologies}},
  author={Ma, Zheng and Xiao, Ming and Xiao, Yue and Pang, Zhibo and Poor, H Vincent and Vucetic, Branka},
  journal={IEEE Internet Things J.},
  volume={6},
  number={5},
  pages={7946--7970},
  year={2019},
  month={Oct.},
  publisher={IEEE}
}

@article{salmeno2024reliable,
  title={{Reliable Beam Tracking on High-Altitude Platform for Millimeter Wave High-speed Railway}},
  author={Salmeno, Amran P and Zakia, Irma},
  journal={IEEE Access},
  year={2024},
  month={Mar.},
  volume={12},
  publisher={IEEE}
}

@article{tang2022deep,
  title={{Deep Reinforcement Learning-Based Resource Allocation for Satellite Internet of Things with Diverse QoS Guarantee}},
  author={Tang, Siqi and Pan, Zhisong and Hu, Guyu and Wu, Yang and Li, Yunbo},
  journal={Sensors},
  volume={22},
  number={8},
  pages={2979},
  year={2022},
  month={Apr.},
  publisher={MDPI}
}

@inproceedings{li2024bistatic,
  title={{A Bistatic Sensing System in Space-Air-Ground Integrated Networks}},
  author={Li, Xiangyu and Shang, Bodong and Wu, Qingqing},
  booktitle={Proc. IEEE/CIC Int. Conf. Commun. China (ICCC)},
  pages={1823-1827},
  year={2024},
  month={Aug.},
}

@article{xu2025robust,
  title={{Robust Secure Beamforming Design for Multi-RIS-aided MISO systems with Hardware Impairments and Channel Uncertainties}},
  author={Xu, Yongjun and Tian, Qinyu and Chen, Qianbin and Wu, Qingqing and Huang, Chongwen and Zhang, Haijun and Yuen, Chau},
  journal={IEEE Trans. Commun.},
  year={2025},
  month={Mar.},
  volume={73},
  number={3},
  publisher={IEEE}
}

@article{jiang2025hybrid,
  title={{A Hybrid Framework of RIS-assisted Robust Secure Transmission Design for Multibeam Satellite Communications}},
  author={Jiang, Chengjun and Zhang, Chensi and Huang, Chongwen and Zhang, Jian and Zhu, Xiaojie and Ge, Jianhua and Debbah, Merouane and Yuen, Chau},
  journal={IEEE Trans. Veh. Technol.},
  year={2025},
  month={Apr.},
  volume={74},
  number={4},
  publisher={IEEE}
}

@inproceedings{dai2024secrecy,
  title={{Secrecy Rate Optimization for Active-Ris Assisted LEO Satellite Communications}},
  author={Dai, Cui-Qin and Wang, Kailong and Liao, Mingxia},
  booktitle={Proc. IEEE 24th Int. Conf. Commun. Technol. (ICCT)},
  pages={1756--1760},
  year={2024},
  month={Oct.},
  organization={IEEE}
}

@article{ibrahim2023joint,
  title={{Joint beamforming optimization design and performance evaluation of RIS-aided wireless networks: A comprehensive state-of-the-art review}},
  author={Ibrahim, Laith and Mahmud, Mohd Nazri and Salleh, Mohd Fadzli Mohd and Al-Rimawi, Ashraf},
  journal={IEEE Access},
  volume={11},
  pages={141801--141859},
  year={2023},
  month={Dec.},
  publisher={IEEE}
}

@article{liu2021reconfigurable,
  title={{Reconfigurable Intelligent Surfaces: Principles and Opportunities}},
  author={Liu, Yuanwei and Liu, Xiao and Mu, Xidong and Hou, Tianwei and Xu, Jiaqi and Di Renzo, Marco and Al-Dhahir, Naofal},
  journal={IEEE Commun. Surveys Tuts.},
  volume={23},
  number={3},
  pages={1546--1577},
  year={2021},
  month={3rd Quart.,},
  publisher={IEEE}
}

@article{li2025downlink,
  title={Downlink Performance of Cell-Free Massive {MIMO} for {LEO} Satellite Mega-Constellation},
  author={Li, Xiangyu and Shang, Bodong},
  journal={IEEE Trans. Mobile Comput.},
  volume={},
  number={},
  pages={1--1},
  year={2025},
  month={Nov.},
  publisher={IEEE}
}

@article{voicu2024handover,
  title={{Handover Strategies for Emerging LEO, MEO, and HEO Satellite Networks}},
  author={Voicu, Andra M and Bhattacharya, Abhipshito and Petrova, Marina},
  journal={IEEE Access},
  year={2024},
  month={Feb.},
  volume={12},
  number={},
  publisher={IEEE}
}

@article{radhakrishnan2016survey,
  title={{Survey of Inter-Satellite Communication for Small Satellite Systems: Physical Layer to Network Layer View}},
  author={Radhakrishnan, Radhika and Edmonson, William W and Afghah, Fatemeh and Rodriguez-Osorio, Ramon Martinez and Pinto, Frank and Burleigh, Scott C},
  journal={IEEE Commun. Surveys Tuts.},
  volume={18},
  number={4},
  pages={2442--2473},
  year={2016},
  publisher={IEEE}
}

@article{prol2022position,
  title={{Position, navigation, and timing (PNT) through low earth orbit (LEO) satellites: A survey on current status, challenges, and opportunities}},
  author={Prol, Fabricio S and Ferre, R Morales and Saleem, Zainab and V{\"a}lisuo, Petri and Pinell, Christina and Lohan, Elena Simona and Elsanhoury, Mahmoud and Elmusrati, Mohammed and Islam, Saiful and {\c{C}}elikbilek, Kaan and others},
  journal={IEEE Access},
  volume={10},
  pages={83971--84002},
  year={2022},
  publisher={IEEE}
}

@article{zheng2024cooperative,
  title={{Cooperative Multi-Satellite and Multi-RIS Beamforming: Enhancing LEO SatCom and Mitigating LEO-GEO Intersystem Interference}},
  author={Zheng, Ziyuan and Jing, Wenpeng and Lu, Zhaoming and Wu, Qingqing and Zhang, Haijun and Gesbert, David},
  journal={IEEE J. Sel. Areas Commun.},
  year={2024},
  volume={43},
  number={1},
  pages={279-296},
  publisher={IEEE}
}

@article{khoshafa2025ris,
  title={{RIS-Empowered Secured Space-Air-Ground Integrated Networks: Opportunities and Challenges}},
  author={Khoshafa, Majid H and Bueno, Felipe and Ngatched, Telex MN and Di Renzo, Marco},
  journal={IEEE Commun. Mag.},
  year={2025},
  volume={63},
  number={6},
  pages={130-136},
  publisher={IEEE}
}

@article{zhang2024generative,
  title={{Generative AI for Space-Air-Ground Integrated Networks}},
  author={Zhang, Ruichen and Du, Hongyang and Niyato, Dusit and Kang, Jiawen and Xiong, Zehui and Jamalipour, Abbas and Zhang, Ping and Kim, Dong In},
  journal={IEEE Wireless Commun.},
  year={2024},
  volume={31},
  number={6},
  pages={10-20},
  publisher={IEEE}
}

@article{khoramnejad2025generative,
  title={{Generative AI for the Optimization of Next-Generation Wireless Networks: Basics, State-of-the-Art, and Open Challenges}},
  author={Khoramnejad, Fahime and Hossain, Ekram},
  journal={IEEE Commun. Surveys Tuts.},
  year={2025},
  volume={},
  number={},
  pages={1-1},
  publisher={IEEE}
}

@article{liu2021compact,
  title={{Compact User-Specific Reconfigurable Intelligent Surfaces for Uplink Transmission}},
  author={Liu, Kunzan and Zhang, Zijian and Dai, Linglong and Hanzo, Lajos},
  journal={IEEE Trans. Commun.},
  volume={70},
  number={1},
  pages={680--692},
  year={2021},
  publisher={IEEE}
}

@article{hu2023securing,
  title={{Securing Uplink Transmissions in Cellular Network via User-Specific Reconfigurable Intelligent Surfaces}},
  author={Hu, Xinyue and Yi, Yibo and Kai, Caihong and Zhang, Yu and Li, Kun},
  journal={IEEE Commun. Lett.},
  volume={27},
  number={9},
  pages={2298--2302},
  year={2023},
  publisher={IEEE}
}

@article{chepuri2023integrated,
  title={{Integrated Sensing and Communications with Reconfigurable Intelligent Surfaces: From Signal Modeling to Processing}},
  author={Chepuri, Sundeep Prabhakar and Shlezinger, Nir and Liu, Fan and Alexandropoulos, George C and Buzzi, Stefano and Eldar, Yonina C},
  journal={IEEE Signal Process. Mag.},
  volume={40},
  number={6},
  pages={41--62},
  year={2023},
  publisher={IEEE}
}

@article{wang2023stars,
  title={{STARS Enabled Integrated Sensing and Communications}},
  author={Wang, Zhaolin and Mu, Xidong and Liu, Yuanwei},
  journal={IEEE Trans. Wireless Commun.},
  volume={22},
  number={10},
  pages={6750--6765},
  year={2023},
  publisher={IEEE}
}

@article{ge2024ris,
  title={{RIS-Assisted Cooperative Spectrum Sensing for Cognitive Radio Networks}},
  author={Ge, Jungang and Liang, Ying-Chang and Wang, Shuo and Sun, Chen},
  journal={IEEE Trans. Wireless Commun.},
  year={2024},
  publisher={IEEE}
}

@article{tekbiyik2021energy,
  title={{Energy-Efficient RIS-Assisted Satellites for IoT Networks}},
  author={Tekb{\i}y{\i}k, K{\"u}r{\c{s}}at and Kurt, G{\"u}ne{\c{s}} Karabulut and Yanikomeroglu, Halim},
  journal={IEEE Internet Things J.},
  volume={9},
  number={16},
  pages={14891--14899},
  year={2021},
  publisher={IEEE}
}

@article{lukito2024integrated,
  title={{Integrated STAR-RIS and UAV for Satellite IoT Communications: An Energy-Efficient Approach}},
  author={Lukito, William D and Xiang, Wei and Lai, Phu and Cheng, Peng and Liu, Chang and Yu, Kan and Zhu, Xiaoyan},
  journal={IEEE Internet Things J.},
  year={2024},
  publisher={IEEE}
}

@article{zou2023machine,
  title={{Machine learning in RIS-assisted NOMA IoT networks}},
  author={Zou, Yixuan and Liu, Yuanwei and Mu, Xidong and Zhang, Xingqi and Liu, Yue and Yuen, Chau},
  journal={IEEE Internet Things J.},
  volume={10},
  number={22},
  pages={19427--19440},
  year={2023},
  publisher={IEEE}
}

@article{marinello2025grant,
  title={{Grant-Free Random Access for RIS-Aided Machine-Type Communication}},
  author={Marinello Filho, Jos{\'e} Carlos and Abr{\~a}o, Taufik and Hossain, Ekram and Mezghani, Amine},
  journal={IEEE Trans. Wireless Commun.},
  year={2025},
  publisher={IEEE}
}

@article{zheng2023leo,
  title={{LEO satellite and RIS: Two Keys to Seamless Indoor and Outdoor Localization}},
  author={Zheng, Pinjun and Liu, Xing and He, Jiguang and Seco-Granados, Gonzalo and Al-Naffouri, Tareq Y},
  journal={arXiv preprint arXiv:2312.16946},
  year={2023}
}

@inproceedings{wang2024beamforming,
  title={{Beamforming Design and Performance Evaluation for RIS-Aided Localization Using LEO Satellite Signals}},
  author={Wang, Lei and Zheng, Pinjun and Liu, Xing and Ballal, Tarig and Al-Naffouri, Tareq Y},
  booktitle={Proc. IEEE Int. Conf. Acoust., Speech Signal Process. (ICASSP)},
  pages={13166--13170},
  year={2024},
}

@article{dureppagari2023ntn,
  title={{NTN-Based 6G Localization: Vision, Role of LEOs, and Open Problems}},
  author={Dureppagari, Harish K and Saha, Chiranjib and Dhillon, Harpreet S and Buehrer, R Michael},
  journal={IEEE Wireless Commun.},
  volume={30},
  number={6},
  pages={44--51},
  year={2023},
  publisher={IEEE}
}

@article{saleh20246g,
  title={{6G RIS-Aided Single-LEO Localization with Slow and Fast Doppler Effects}},
  author={Saleh, Sharief and Keskin, Musa Furkan and Priyanto, Basuki and Beale, Martin and Zheng, Pinjun and Al-Naffouri, Tareq Y and Seco-Granados, Gonzalo and Wymeersch, Henk},
  journal={arXiv preprint arXiv:2410.11010},
  year={2024}
}

@ARTICLE{Wref65,
  author={He, Hongmei and Zhou, Di and Sheng, Min and Li, Jiandong},
  journal={IEEE Trans. Commun.}, 
  title={{Hierarchical Cross-Domain Satellite Resource Management: An Intelligent Collaboration Perspective}}, 
  year={2023},
  volume={71},
  number={4},
  pages={2201-2215},
}

@ARTICLE{Wref67,
  author={Kunz, Andreas and Baskaran, Sheeba Backia Mary and Alexandropoulos, George C.},
  journal={IEEE Commun. Standards Mag.}, 
  title={{Lightweight Security for Ambient-Powered Programmable Reflections with Reconfigurable Intelligent Surfaces}}, 
  year={2025},
  volume={},
  number={},
  pages={1-1},
}

@ARTICLE{Wref68,
  author={Bannour, Fetia and Souihi, Sami and Mellouk, Abdelhamid},
  journal={IEEE Commun. Surveys Tuts.}, 
  title={{Distributed SDN Control: Survey, Taxonomy, and Challenges}}, 
  year={2018},
  volume={20},
  number={1},
  pages={333-354},
}

@article{WrefAI,
  author    = {Khoramnejad, F. and Hossain, E.},
  title     = {{Generative AI for the Optimization of Next-Generation Wireless Networks: Basics, State-of-the-Art, and Open Challenges}},
  journal   = {IEEE Commun. Surveys Tuts.},
  pages     = {1--1},
  year      = {2025}
}

@ARTICLE{Wref70,
  author={Khoramnejad, Fahime and Hossain, Ekram},
  journal={IEEE Commun. Surveys Tuts.}, 
  title={{Generative AI for the Optimization of Next-Generation Wireless Networks: Basics, State-of-the-Art, and Open Challenges}}, 
  year={2025},
  volume={},
  number={},
  pages={1-1}
}

@ARTICLE{Wref71,
  author={Thomas, Anna and Deka, Kuntal and Sharma, Sanjeev and Rajamohan, Neelakandan},
  journal={IEEE Trans. Veh. Technol.}, 
  title={{IRS-Assisted OTFS System: Design and Analysis}}, 
  year={2023},
  volume={72},
  number={3},
  pages={3345-3358},
}

@INPROCEEDINGS{Zheng_deployment,
  author={Zheng, Ziyuan and Jing, Wenpeng and Lu, Zhaoming and Wen, Xiangming},
  booktitle={Proc. IEEE Globecom Workshops (GC Wkshps)}, 
  title={{RIS-Enhanced LEO Satellite Communication: Joint Passive Beamforming and Orientation Optimization}}, 
  year={2022},
  volume={},
  number={},
  pages={874-879},
}

@ARTICLE{Zheng_hotspot,
  author={Zheng, Ziyuan and Jing, Wenpeng and Lu, Zhaoming and Wen, Xiangming and Wu, Qingqing and Shao, Hua},
  journal={IEEE Trans. Wireless Commun.}, 
  title={{RIS-Aided Hotspot Capacity Enhancement for Multibeam Satellite Systems}}, 
  year={2024},
  volume={23},
  number={4},
  pages={3648-3664},
}

@INPROCEEDINGS{Zheng_coverage,
  author={Cao, Ning and Zheng, Ziyuan and Jing, Wenpeng and Lu, Zhaoming and Wen, Xiangming},
  booktitle={2023 IEEE 34th Annual International Symposium on Personal, Indoor and Mobile Radio Communications (PIMRC)}, 
  title={{RIS-Assisted Coverage Extension for LEO Satellite Communication in Blockage Scenarios}}, 
  year={2023},
  volume={},
  number={},
  pages={1-6},
}

@ARTICLE{Zheng_SatIRS,
  author={Zheng, Beixiong and Lin, Shaoe and Zhang, Rui},
  journal={IEEE J. Sel. Areas Commun.}, 
  title={{Intelligent Reflecting Surface-Aided LEO Satellite Communication: Cooperative Passive Beamforming and Distributed Channel Estimation}}, 
  year={2022},
  volume={40},
  number={10},
  pages={3057-3070},
}

@ARTICLE{QQIRS,
  author={Wu, Qingqing and Zhang, Rui},
  journal={IEEE Trans. Wireless Commun.}, 
  title={{Intelligent Reflecting Surface Enhanced Wireless Network via Joint Active and Passive Beamforming}}, 
  year={2019},
  volume={18},
  number={11},
  pages={5394-5409},
}

@ARTICLE{AlouiniSat,
  author={Yaacoub, Elias and Alouini, Mohamed-Slim},
  journal={Proc. IEEE}, 
  title={{A Key 6G Challenge and Opportunity—Connecting the Base of the Pyramid: A Survey on Rural Connectivity}}, 
  year={2020},
  volume={108},
  number={4},
  pages={533-582},
}

@article{xv2023joint,
  title={Joint beam scheduling and beamforming design for cooperative positioning in multi-beam LEO satellite networks},
  author={Xv, Hongtao and Sun, Yaohua and Zhao, Yafei and Peng, Mugen and Zhang, Shijie},
  journal={IEEE Trans. Veh. Technol.},
  volume={73},
  number={4},
  pages={5276--5287},
  year={2023},
  publisher={IEEE}
}

@article{kim2024joint,
  title={Joint user scheduling and phase shift optimization for STAR-RIS-assisted multicast satellite communications},
  author={Kim, Taehyoung and Jung, Minchae and Son, Hyukmin},
  journal={IEEE Trans. Aerosp. Electron. Syst.},
  volume={60},
  number={5},
  pages={7466--7474},
  year={2024},
  publisher={IEEE}
}

@inproceedings{hugh2025resource,
  title={Resource-Aware Network Slicing for QoS-Driven Service Orchestration in Integrated Satellite--5G Systems},
  author={Hugh, Q and Soria, Freddy},
  booktitle={ECCSUBMIT Conferences},
  volume={3},
  number={2},
  pages={70--75},
  year={2025}
}

@inproceedings{momani2025beam,
  title={Beam Update Rate Analysis for Low-Complexity Hybrid Beamforming in LEO Satellites},
  author={Momani, Mohammad and Delamotte, Thomas and Knopp, Andreas},
  booktitle={Proc. 12th Adv. Satell. Multimedia Syst. Conf. 18th Signal Process. Space Commun. Workshop (ASMS/SPSC)},
  pages={1--6},
  year={2025},
  organization={IEEE}
}

@article{lin2022refracting,
  title={Refracting RIS-aided hybrid satellite-terrestrial relay networks: Joint beamforming design and optimization},
  author={Lin, Zhi and Niu, Hehao and An, Kang and Wang, Yong and Zheng, Gan and Chatzinotas, Symeon and Hu, Yihua},
  journal={IEEE Trans. Aerosp. Electron. Syst.},
  volume={58},
  number={4},
  pages={3717--3724},
  year={2022},
  publisher={IEEE}
}

@article{morel2015link,
  title={Link adaptation strategies for next generation satellite video broadcasting: A system approach},
  author={Morel, Coline and Arapoglou, Pantelis-Daniel and Angelone, Martina and Ginesi, Alberto},
  journal={IEEE Trans. Broadcast.},
  volume={61},
  number={4},
  pages={603--614},
  year={2015},
  publisher={IEEE}
}

@article{liu2025unintended,
  title={Unintended Interference Suppression Based on Decision Feedback Adaptive Cancellation for DSSS Satellite Communication},
  author={Liu, Jiancheng and Zhang, Jingtao and Li, Tian and Zhou, Yu and Wang, Yanpeng},
  journal={Chin. J. Electron.},
  volume={34},
  number={2},
  pages={673--682},
  year={2025},
  publisher={CIE}
}

@article{li2025enriched,
  title={Enriched K-Tier Heterogeneous Satellite Networks Model With User Association Policies}, 
  author={Li, Zhuhang and Shang, Bodong},
  journal={IEEE Trans. Veh. Technol.}, 
  year={2025},
  volume={},
  number={},
  pages={1-16},
  publisher={IEEE}
}

@article{abudureheman2025resource,
  title={Resource allocation in multi-beam leo satellite systems based on beam hopping and frequency reuse},
  author={Abudureheman, Yasenjiang and Chu, Jianxiang and Song, Ruoxi and Liu, Xiangnan and Huangfu, Wei and Zhang, Haijun},
  journal={IEEE Internet Things J.},
  year={2025},
  publisher={IEEE}
}

@article{han2025demand,
  title={On-demand optimization method for cross-layer topology in multi-task VLEO and mega-LEO heterogeneous satellite networks},
  author={Han, Kai and Siew, Marie and Xu, Bingbing and Guo, Shengjun and Gong, Wenbin and Quek, Tony QS and Ren, Qianyi},
  journal={IEEE Trans. Wireless Commun.},
  year={2025},
  volume={24},
  number={11},
  pages={9598-9612},
  publisher={IEEE}
}

@article{park2022tractable,
  title={A tractable approach to coverage analysis in downlink satellite networks},
  author={Park, Jeonghun and Choi, Jinseok and Lee, Namyoon},
  journal={IEEE Trans. Wireless Commun.},
  volume={22},
  number={2},
  pages={793--807},
  year={2022},
  month={Aug.},
  publisher={IEEE}
}

@article{okati2023stochastic,
  title={Stochastic coverage analysis for multi-altitude LEO satellite networks},
  author={Okati, Niloofar and Riihonen, Taneli},
  journal={IEEE Commun. Lett.},
  volume={27},
  number={12},
  pages={3305--3309},
  year={2023},
  publisher={IEEE}
}

@inproceedings{Hadani2017OTFSWCNC,
  author    = {R. Hadani and S. Rakib and M. Tsatsanis and A. Monk and A. J. Goldsmith and A. F. Molisch and R. Calderbank},
  title     = {Orthogonal Time Frequency Space Modulation},
  booktitle = {Proc. IEEE Wireless Communications and Networking Conference (WCNC)},
  year      = {2017},
}

@article{Raviteja2018TWC_OTFS_IC,
  author  = {P. Raviteja and K. T. Phan and Y. Hong and E. Viterbo},
  title   = {Interference Cancellation and Iterative Detection for Orthogonal Time Frequency Space Modulation},
  journal = {IEEE Transactions on Wireless Communications},
  year    = {2018},
  volume  = {17},
  number  = {10},
  pages   = {6501--6515},
}

@article{Raviteja2019TVT_OTFS_CE,
  author  = {P. Raviteja and K. T. Phan and Y. Hong},
  title   = {Embedded Pilot-Aided Channel Estimation for OTFS in Delay--Doppler Channels},
  journal = {IEEE Transactions on Vehicular Technology},
  year    = {2019},
  volume  = {68},
  number  = {5},
  pages   = {4906--4917},
}

@article{Bhat2023CL_RISOTFS_Fractional,
  author  = {V. S. Bhat and G. Harshavardhan and A. Chockalingam},
  title   = {Input-Output Relation and Performance of RIS-Aided OTFS With Fractional Delay--Doppler},
  journal = {IEEE Communications Letters},
  year    = {2023},
  volume  = {27},
  number  = {1},
  pages   = {337--341},
}

@article{Tao2025OJVT_RIS_OTFS_Survey,
  author  = {Q. Tao and Z. Li and K. Zhi and S. Li and W. Yuan and L. Zaniboni and S. Stanczak and E. Viterbo and X. Wang},
  title   = {A Survey on Reconfigurable Intelligent Surface-Assisted Orthogonal Time Frequency Space Systems},
  journal = {IEEE Open Journal of Vehicular Technology},
  year    = {2025},
  volume  = {6},
  pages   = {1881--1909},
}

@article{Shen2022JSAC_MassiveMIMOOTFS,
  author  = {B. Shen and Y. Wu and J. An and C. Xing and L. C. Zhao and W. Zhang},
  title   = {Random Access With Massive MIMO-OTFS in LEO Satellite Communications},
  journal = {IEEE Journal on Selected Areas in Communications},
  year    = {2022},
  volume  = {40},
  number  = {10},
  pages   = {2865--2881},
}

\section*{Author contributions}
Z.Z. led the writing of the manuscript, contributed to sections 1-6, and prepared figures 2-4. 
X.L. contributed to sections 2-5 and prepared figure 1.
S.Z. contributed to sections 2 and 5.
Y.W. contributed to section 4 and prepared figure 5.
Q.W. conceived the original idea.
Q.W., B.S, and W.J. suggested the outline of the article and provided advice and supervision.
All authors reviewed and agreed on the manuscript before submission.

\section*{Competing interests}
All authors declare no financial or non-financial competing interests.
 
\section*{Acknowledgements}
This study was funded by the National Key R\&D Program of China under grant 2022YFB2903500, and the National Natural Science Foundation of China under grants 62371289 and 62331022; 
in part by the Zhejiang Provincial Natural Science Foundation of China under grant LQN25F010003; 
in part by the State Key Laboratory of Integrated Services Networks, Xidian University, China;
in part by the Xiongan New Area Science and Technology Innovation Program grant 2023XAGG0091.

\end{document}